# *In Situ* Epitaxy of Pure Phase Ultra-Thin InAs-Al Nanowires for Quantum Devices

Dong Pan,[†,§,⊥] Huading Song,[‡,§] Shan Zhang,[‡] Lei Liu,[†] Lianjun Wen,[†] Dunyuan Liao,[†] Ran Zhuo,[†] Zhichuan Wang,[△] Zitong Zhang,[‡] Shuai Yang,[‡,§] Jianghua Ying,[‡,§] Wentao Miao,[‡] Yongqing Li,[△] Runan Shang,[§] Hao Zhang,[*,‡,§,∥] and Jianhua Zhao[*,†,§,⊥]

[†]*State Key Laboratory of Superlattices and Microstructures, Institute of Semiconductors, Chinese Academy of Sciences, P.O. Box 912, Beijing 100083, China*
[‡]*State Key Laboratory of Low Dimensional Quantum Physics, Department of Physics, Tsinghua University, 100084 Beijing, China.*
[§]*Beijing Academy of Quantum Information Sciences, 100193 Beijing, China.*
[△]*Beijing National Laboratory for Condensed Matter Physics, Institute of Physics, Chinese Academy of Sciences, Beijing 100190, China*
[∥]*Frontier Science Center for Quantum Information, 100084 Beijing, China*
[⊥]*Center of Materials Science and Optoelectronics Engineering, University of Chinese Academy of Sciences, Beijing, 100190, China*
(Dated: November 27, 2020)

**ABSTRACT:**
Hybrid semiconductor-superconductor InAs-Al nanowires with uniform and defect-free crystal interfaces are one of the most promising candidates used in the quest for Majorana zero modes (MZMs). However, InAs nanowires often exhibit a high density of randomly distributed twin defects and stacking faults, which result in an uncontrolled and non-uniform InAs-Al interface. Furthermore, this type of disorder can create potential inhomogeneity in the wire, destroy the topological gap, and form trivial sub-gap states mimicking MZM in transport experiments. Further study shows that reducing the InAs nanowire diameter from growth can significantly suppress the formation of these defects and stacking faults. Here, we demonstrate the *in situ* growth of ultra-thin InAs nanowires with epitaxial Al film by molecular-beam epitaxy. Our InAs diameter (~ 30 nm) is only one-third of the diameters (~ 100 nm) commonly used in literatures. The ultra-thin InAs nanowires are pure phase crystals for various different growth directions, suggesting a low level of disorder. Transmission electron microscopy confirms an atomically sharp and uniform interface between the Al shell and the InAs wire. Quantum transport study on these devices resolves a hard induced superconducting gap and 2*e*- periodic Coulomb blockade at zero magnetic field, a necessary step for future MZM experiments. A large zero bias conductance peak with a peak height reaching 80% of $2e^2/h$ is observed.

*To whom correspondence should be addressed. E-mails: jhzhao@semi.ac.cn (J.H.Z.); hzquantum@mail.tsinghua.edu.cn (H.Z.)

**KEYWORDS:** *Semiconductor-superconductor heterostructures; ultra-thin nanowires; pure phase; hard superconducting gap; zero bias peak; 2e Coulomb blockade*

Semiconductor nanowires proximately coupled to superconductors have attracted increasing interest in recent years due to the prospect of hosting topologically protected Majorana zero modes (MZMs), which can be used in fault-tolerant quantum computing.[1-9] Among them, InAs-Al and InSb-Al nanowires are of particular interest due to their strong spin-orbit coupling and large Zeeman splitting.[10-16] Many experimental signatures for possible MZMs in structures based on these hybrid nanowires have been reported.[10-14] These experimental studies and related theoretical work have demonstrated that an atomic scale uniform and defect-free crystal interface between the semiconductor and the superconductor is a crucial ingredient for high quality experiment.[17-19,15,20] This interface quality mainly depends on two factors. The first is related to the quality of the superconductor layer. The superconductor layer fabricated by conventional (*ex situ* evaporation/sputter) methods often show undesired low-energy states within the proximity-induced superconducting gap: the soft gap problem. The soft gap is a source of decoherence and thus detrimental to future topological quantum information processing. This problem is solved by Krogstrup *et al.* using *in situ* deposition of an Al shell on the sidewalls of InAs nanowires by molecular-beam epitaxy (MBE).[19] To date, high-quality Al layers have been successfully grown on the InAs, InSb and InAsSb nanowires,[19,16,20-23] see also related networks[24-26] using this low temperature epitaxial technique. The other factor is related to the quality of the semiconductor nanowires. In particular, InAs nanowires usually exhibit random mixtures of wurtzite (WZ) and zinc-blende (ZB) crystal structures, which deteriorates the electrical transport properties due to electron scattering at stacking faults or twin planes.[27,28] For example, this type of scattering related disorder can easily create unwanted quantum dots which lead to the formation of Andreev bound states, the most popular alternative (trivial) explanation of current MZM experimental signatures.[29-31] Theoretical and experimental studies demonstrated that controlling the diameter down to a small value is one of the effective methods to obtain an excellent crystal phase purity for InAs nanowires.[32-35] Nevertheless, to the best of our knowledge, all the InAs-Al nanowires reported have an InAs diameter larger than ~ 40 nm[19-21,36] (typically ~ 100 nm) which often show non-single crystal structure. The epitaxy of ultra-thin InAs-Al nanowires (with pure phase InAs) for quantum devices have not been experimentally demonstrated yet.

Besides the material quality, another motivation for the pursuit of thin diameter is aiming to reduce the number of occupied sub-bands in the wire. Previous experiments likely have multi sub-bands occupied with the precise occupation number unknown. In a multi-band system, each spin-resolved sub-band could host a pair of MZMs, leading to multiple MZMs overlapping in the wire's ends. The coupling between these spatially overlapped MZMs can be significantly suppressed by the smooth potential variation in the system. In transport experiment, they could mimic the Majorana signature, e.g. a quantized zero bias conductance peak, and thus are dubbed as quasi-MZMs with a topological trivial origin.[37,38] Another disadvantage of multi-band is that the tunnel coupling between the probing lead and the proximitized wire is dominated by the lower sub-bands. However, it is the top sub-band which hosts MZMs. The small tunnel coupling between the top sub-band and the probe could significantly suppress the MZM signal (e.g. the MZM-peak width and the peak height after thermal averaging). To reduce the number of occupied sub-bands, electrostatic gating, however, may have a limited effect due to the superconductor screening based on recent Schrodinger-Poisson simulations.[39-41] Therefore, one possible solution is to reduce the wire diameter which increases the sub-band energy spacing.

Here, we present the epitaxy of InAs-Al nanowires with the InAs diameter significantly reduced. The InAs is pure phase crystal for various different growth directions due to its thin diameter. This diameter (~ 30 nm) is only one-third of those used in literatures (typically ~ 100 nm). The ratio of 1:3 in diameter translates into an energy ratio of sub-band spacing of 9:1. This large sub-band spacing may help to reduce the number of occupied sub-bands or even reach the single sub-band regime, a true one-dimensional electron system. For the Al deposition, we found that the substrate temperature is a crucial parameter. By optimizing these parameters, we can obtain continuous and smooth Al shells on the ultra-thin InAs nanowires. Transmission electron microscope (TEM) study confirms that our InAs-Al interface is atomically sharp and uniform. As a direct consequence, these hybrid-wires show hard induced superconducting gap and 2$e$- periodic Coulomb blockade in the electron transport study of corresponding quantum devices. Large zero bias peaks with a peak height reaching 80% of $2e^2/h$ are observed. Our result paves the way for the realization of single sub-band MZMs based on ultra-thin hybrid nanowires.

**Morphology control of the InAs-Al nanostructures.** We start by growing ultra-thin InAs nanowires on Si (111) substrates using Ag as seed particles. The InAs nanowires with diameter ranging from ~20 to 30 nm were obtained by evaporating a thin Ag layer to generate small Ag nanoparticles. We find that the substrate temperature plays a key role in the evolution of the Al shell morphology: a low substrate temperature is a prerequisite in enabling the deposition of continuous Al layers on the nanowire facets. **Figure 1**a,b shows the scanning electron microscope (SEM) images of the InAs-Al recorded from different regions on the Si (111) substrates with the Al shell grown at a relatively high temperature of ~ 1 °C. The Al shells are discontinuous and look like 'pearls on a string' on the side walls of all the InAs nanowires. Figure 1c-e is a close-up of such a typical InAs-Al where the Al crystalline islands have a length distribution ranging from a few nanometers to about several tens nanometers. These crystals are formed due to a high Al adatom mobility and a long adatom diffusion length since the substrate temperature was relative high during growth. This phenomenon was also observed in the growth of Al on thick InAs nanowires by other groups.[19,20,36] The interface between InAs nanowire and Al island is relative sharp since the InAs nanowires are pure phase crystals, free of stacking faults and twin defects, which can provide smooth InAs side surfaces for superconductor growth. To obtain continuous Al shells, we further cooled down the substrate to ~ -10 °C for Al growth. Figure 1f,g shows the corresponding SEM images from different regions on the Si (111) substrates. Now, the Al shells are continuously formed on the InAs nanowire facets, covering the wire's entire length, see Figure 1h-j for a close-up of a typical InAs-Al wire. We note that all the Al shells have a rough and faceted outer surface. The grain structure and faceted surface are further confirmed by the TEM analysis (Section 1, Figure S1 in the Supporting Information). One reason for this rough surface could be that the -10 °C substrate temperature is still not low enough and the Al growth follows the high temperature dynamics. That is, the large Al grains were initially nucleated and well separated. Then new preferred crystal orientations appeared at a later stage of the growth as the role of the grain boundaries and strain contributions. Thus, polycrystalline Al shells which consist of type-$\alpha$ and -$\beta$ grains can be observed by TEM (see Figure S2 in the Supporting Information). Another reason could be that the thickness of Al shell measured from HRTEM images is ~ 30 nm, which is beyond the critical thickness of Al layer for smooth shell growth. In short, the substrate temperature is a crucial parameter in enabling the deposition of continuous Al layers on the ultrathin InAs nanowires and the Al thickness directly affects the smoothness of the epitaxial Al layers.

**Ultra-thin InAs-Al nanowires grown under the optimum conditions.** Based on the feedback above, we further lower the substrate temperature during Al growth and reduce the Al film thickness (less than a critical thickness ~ 15 nm in our case). Thin Al shell is in any case necessary for its superconductivity to be maintained at high magnetic field. **Figure 2**a is the SEM image of an ultra-thin InAs-Al nanowires grown under this 'optimum conditions': the growth temperature and growth time of the Al shells are ~-40 °C and 3 min, respectively. It is evident from Figure 2a that continuous and smooth Al half-shells are formed on the facets of almost all the InAs nanowires. Figure 2c-h shows the close-up SEM images of the ultra-thin InAs-Al nanowires. Notably, the diameters of the InAs-Al nanowires are very small: ranging from ~35 nm to ~45 nm. We can also see from the Figure 2b (enlarge view of Figure 2a) that some InAs nanowires have a natural shadowed region which separates the Al shell into two islands. The special morphology features of this kind of nanowires can be used to estimate the thickness of the Al shells. As shown in Figure 2h, the thickness of the shadowed region (pure InAs nanowire) and the segment covered with Al is ~18 nm and ~35 nm, respectively. Thus, the thickness of the Al shell (including native oxide layer) is ~ 17 nm, matching our expectations based on growth rate and time. Thinner Al films (down to ~ 6 nm thickness) can also be achieved using this method (not shown here). Detailed SEM measurements suggest that continuous, uniform and thin half Al shells have been successfully grown on the ultrathin InAs nanowire facets (Section 2, Supporting Information).

**Crystal structure and chemical composition of the ultra-thin InAs-Al nanowires.** We next perform TEM and energy dispersive spectrum (EDS) measurements on our ultra-thin wires to evaluate the InAs-Al interface quality, which is crucial for MZM experiments. **Figure 3**a is a typical low-magnification TEM image showing a continuous and smooth half Al shell on the facet of the InAs nanowires, consistent with the SEM results in Figure 2. Figure 3b is a high-resolution TEM image viewed along the [0001] axis. The Al shell is well epitaxially grown on the InAs nanowire with no dislocations in both the InAs and the Al phases. Meanwhile, we do not have axial and radial rotations of the Al which happens in the case of InAsSb-Al[23]. According to the FFTs of the Al (Figure 3d) and InAs (Figure 3e) segments viewed along the [0001] axis, the InAs nanowire has a pure WZ crystal structure and the Al has a pure ZB crystal structure. Figure 3c shows a high-resolution TEM image of the InAs-Al nanowire viewed along the [2-1-10] axis. We can see that the Al shell forms a perfectly sharp and uniform interface to the InAs nanowire. The thicknesses of the InAs nanowire and Al shell are ~ 19 nm and ~ 15 nm, respectively. The FFT of the InAs (Figure 3f) segment viewed along the [2-1-10] axis further confirms that InAs nanowire has a pure WZ crystal structure with no stacking faults or twin defects. The interface quality of the ultra-thin InAs-Al nanowire also has been identified by high-angle annular darkfield scanning transmission electron microscopy (HAADF-STEM) and EDS measurements. Figure 3g,h is HAADF-STEM images taken from the top and middle sections of the ultra-thin InAs-Al nanowire, respectively. This again shows that the ultra-thin InAs-Al nanowire has a very sharp and uniform interface. The false-color EDS elemental maps of In (Figure 3i), As (Figure 3j), Al (Figure 3k) and their overlay (Figure 3l) taken at the middle region of the InAs-Al nanowire further confirm that a very sharp and uniform interface formed between the InAs nanowire and the Al layer. Detailed TEM observation of a dozen such InAs-Al nanowires can be found in Supporting Information (Section 3), all revealed an atomic scale uniformity of the interface and a perfect crystal structure (either pure ZB or pure WZ), free of stacking faults. Continuous, uniform and thin half Al shells can be well epitaxially grown on the facets of InAs nanowires with all the growth directions (see Figure S6-S11 in the Supporting Information).

**Hard superconducting gap.** We now turn to the quantum transport study of our ultra-thin InAs-Al devices. **Figure 4**a shows an normal-nanowire-superconductor (N-NW-S) device (device A), where the global back gate ($V_{BG}$) tunes the electron density in the InAs nanowire (grey), as well as the tunnel barrier (the wire segment with Al shell etched away). Details of device fabrication and measurement circuit can be found in the Experimental Section. The InAs diameter in this device is ~ 34 nm. Figure 4b shows the differential conductance (d$I$/d$V$) as a function of the source-drain bias voltage ($V$) and $V_{BG}$. Sweeping $V_{BG}$ more negative, the overall conductance decreases to zero, tuning the barrier segment of the nanowire from open to pinched off. The two horizontal white lines features (around ±0.24 mV) indicate the superconducting gap edge. In the tunneling regime (near pinched off), d$I$/d$V$ reveals the density of states in the proximitized nanowire part, as shown in Figure 4c-d. The sub-gap conductance is suppressed to zero, suggesting a hard induced superconducting gap.[15,16,19,42,43] This hard gap is a direct result from the high-quality atomic-flat InAs-Al interface as demonstrated in Figure 3.[44] The suppression ratio of the sub-gap conductance compared with the normal state conductance (outside gap) is larger than 50, comparable with the values reported in literature.[15,16,19,42,43] The superconducting gap remains hard with sub-gap conductance sticking to zero after applying a magnetic field ($B$), as shown in Figure 4e. The $B$-direction is along the wire axis. The gap closes around 1.1 Tesla, due to orbital effect of the bulk Al film with an estimated film thickness of 15 nm.

**Large zero bias peak.** At a different $V_{BG}$, the d$I$/d$V$ spectroscopy of Device A reveals an Andreev bound state indicated by the two symmetric sub-gap peaks at $B = 0$ as shown in **Figure 5**a. We note several charge rearrangements between the measurement of Figure 4 and Figure 5, which re-set the device condition. The two peaks in Figure 5a Zeeman split in $B$-field. The inner peaks merge at zero bias, forming a zero bias peak (ZBP). The effective g-factor is estimated to be ~ 7.6 based on the Zeeman splitting dispersion in $B$.[24,45] This ZBP remains non-split against $B$-sweep from 0.28 T to 1.1 T. This $B$-range of ~ 0.8 T translates into a Zeeman energy scale (1/2g$\mu_B B$) of ~0.18 meV, comparable with the superconducting gap and much larger than the ZBP-width. A ZBP sticking at zero bias in B-sweep was initially regarded as a possible sign of MZM in hybrid nanowire devices.[10,46,47] However, a more likely alternative explanation with a trivial origin based on Andreev bound states was quickly proposed.[48-50,31] Figure 5b shows the zero-bias line-cut where the ZBP-height first reaches 80% of 2$e^2$/$h$, and then decreases by further increasing $B$-field. We note that our reported d$I$/d$V$ is purely the conductance of the device (including unknown contact resistance) with the outer measurement circuit resistance (e.g. fridge filters) subtracted based on independent calibration (see Experimental section for details). Figure 5c and Figure 5d show the d$I$/d$V$ curve at 0 T and 0.39 T, respectively. Our ZBP height is close to 2$e^2$/$h$ and significantly larger than the ZBP-height reported on similar InAs-Al devices in literature.[46] The relation between ZBP-height and wire diameter can be studied in future research.[51,52]. Figure 5e shows another ZBP at a different $V_{BG}$ which also remains non-split over a sizable $B$-range. This ZBP-height, however, is much less than the one in Figure 5a. Figure 5f shows the $V_{BG}$ sweep at 0.5 T which reveals the two ZBPs in Figure 5a and e as two level-crossing points, similar to the typical Andreev bound state peaks reported in literature.[48] This level-crossing feature of ZBP rules out its topological origin[45] since MZM induced ZBPs should remain non-split over a sizable range in both $B$-scan and gate-scans. The full phase diagram (in $B$ and $V_{BG}$ space) of this trivial ZBP is shown in Supplement Figure S12 and Figure S13.

**2$e$-periodic Coulomb blockade and 2$e$-1$e$ transition of an island device.** Next we study a different device with the Al shell floated instead of grounded as shown in **Figure 6**a. The InAs wire diameter in this device is ~ 34 nm. The nanowire segment with the floated Al island tunnel couples to the source and drain normal metal contacts through the InAs nanowire, forming a gate-tunable hybrid superconducting island device with a finite Coulomb charging energy. The plunger gate ($V_{PG}$) tunes the electron number on the island, leading to Coulomb blockade diamonds in the d$I$/d$V$ spectroscopy as shown in Figure 6b. The hard induced superconducting gap ensures negligible quasi-particle poisoning. Therefore, each Coulomb diamond in Figure 6b corresponds to 2$e$ (Cooper pair) instead of 1$e$ (quasi-particle) charge transport. This 2$e$ feature is also reflected by comparing the Coulomb oscillation period at zero bias (2$e$) and high bias (1$e$) in Figure 6c. Based on the Coulomb diamond size ($V$ ~ 0.22 meV), we estimate the charging energy of $E_C = e^2/2C$ ~ 28 μeV, smaller than the superconducting gap and thus fulfilling a necessary condition for 2$e$. Figure 6d shows the d$I$/d$V$ curve at the Coulomb blockade valley (red) and the degeneracy point (black). Negative differential conductance (NDC) is observed at the degeneracy point which is known features for these hybrid islands. Based on the onset bias of NDC and $V_{NDC} = 2(E_o - E_C)/e$ ~ 68 μeV,[53] we estimate the energy of the lowest sub-gap state $E_O$ ~ 62 μeV. In Supplement Figure S14, we show another island device (with InAs wire diameter ~ 28 nm) where the extracted parameters are in similar ranges. Figure 6e shows the 2$e$-periodic Coulomb peaks (at $V$ = 0 mV) splitting into two 1$e$-periodic peaks in a $B$-sweep (direction along the nanowire), see Figure 6f typical line-cuts. This 2$e$-1$e$ transition is likely due to the closing of the superconducting gap: when the gap or $E_O$ is smaller (lower) than $E_C$, odd parity ground state starts to be available, and the 2$e$-peak splits. The $B$-field induced 2$e$-1$e$ transition and its further peak-spacing-oscillation pattern is another possible MZM signature in these devices,[54] with also alternative explanations proposed based on trivial origins.[55,56]

In summary, we have developed an MBE growth technique for the epitaxy of high-quality Al layers on the pure phase, ultra-thin InAs nanowires for various different growth directions. A sharp and uniform interface is identified. As a result, electron transport reveals a hard induced superconducting gap and 2e-Coulomb blockade in related quantum devices. Our results provide a material platform for searching MZMs in fewer sub-bands hybrid nanowire quantum devices.

**Experimental Section**

*InAs-Al Nanowire Growth*: InAs nanowires were grown in a solid source MBE system (VG 80) equipped with standard Knudsen-cells. Commercial p-type Si (111) wafers were used as the substrates. Before loading the Si substrates into the MBE chamber, they were immersed in a diluted HF (2%) solution for 1 min to remove the surface contamination and native oxide.[35] The Si (111) substrates were initially heated in the preparation chamber to ~200 °C for water desorption. After that, a Ag layer less than 0.5 nm nominal thickness was deposited on the Si substrate in the MBE growth chamber at room temperature and then annealed *in situ* at 550 $^0$C for 20 min to generate small Ag nano-particles. Ultrathin InAs nanowires with a diameter ranging from ~20 to 30 nm were obtained with these small Ag seed particles (very few nanowires with a diameter smaller than 10 nm or larger than 40 nm also appear on the substrate surface resulting from the Ostwald ripening). According to our previous work,[57] the dimension of InAs grown by MBE using Ag as catalysts can be tuned directly from one-dimensional nanowires to two-dimensional nanosheets and to three-dimensional complex crosses only by increasing the indium flux. Thus, to obtain InAs nanowires, the sample should be grown under an arsenic rich growth condition. In this work, the ultra-thin InAs nanowires were grown for 40-80 min at a temperature of 505 $^0$C with an arsenic/indium beam equivalent pressure ratio of ~ 42 (the beam fluxes of In and As$_4$ sources are $1.1 \times 10^{-7}$ mbar and $4.6 \times 10^{-6}$ mbar, respectively). After the growth of nanowires, the sample is transferred from the growth chamber to the preparation chamber at 300 °C to avoid arsenic condensation on the nanowire surface. The sample is then cooled down to low temperatures (~ 1 °C to -40 °C) by natural cooling and liquid nitrogen cooling. Al is evaporated from a Knudsen cell at an angle of ~20° from the substrate normal (~70° from substrate surface) and at a temperature of ~1150 °C for 180 s-360 s (giving approximately 0.08 nm/s). To obtain half Al shells, the substrate rotation is kept disabled during the Al growth. When the growth of nanowires with Al is completed, the sample is rapidly pulled out of the MBE growth chamber and oxidized naturally.

*Morphological, Structural and Chemical Composition Characterizations*: Morphologies of the InAs-Al samples were investigated by SEM (FEI NanoSEM 650 and FEI Helios G4 CX) operated at 10-15 kV. Crystal structure of the InAs-Al samples was investigated by a JEM F200 TEM operated at 200 kV. Elemental distribution of the InAs-Al samples was measured by EDS based on the STEM mode using the JEM F200 equipped with dual high sensitivity large silicon drift detectors. For TEM characterization, nanowires were removed from the growth substrate via sonication in ethanol and then drop-cast onto lacey carbon grids.

*Device Fabrication*: InAs-Al nanowires were transferred by cleanroom wipes touching and swiping the growth chip, then swiping it onto the chip substrate: highly p-doped Si covered by 285 nm thick silicon dioxide. The aluminum film was selectively etched using Transene Aluminum Etchant Type D at 50 °C for 10 seconds. Ohmic contacts to the InAs nanowire were fabricated by 80 s Argon plasma etching at a power of 50 W and pressure of 0.05 Torr, followed by metal evaporation of Ti/Au (10/70 nm).

*Transport measurement*: The devices were cooled down and measured in a Bluefors dilution refrigerator at a base temperature ~ 20 mK, equipped with a 6-1-1 T vector magnet. Differential conductance was measured using an AC lock-in technique, with excitation voltage of 8 μV. Fridge filter resistance was independently calibrated and subtracted. The measurement circuit was also calibrated with testing resistors. During the measurement of device A, the lock-in notch filter was on, which underestimates the ZBP height (Figure 5a) by ~ 3%. The nanowire orientation was identified by rotation the magnetic field direction in the substrate plane: the superconducting gap is maximum when the field is aligned with the wire. This is also consistent with the direction estimated based on the device optical/SEM image and chip mounting.

ASSOCIATED CONTENT

Supporting Information

AUTHOR INFORMATION

**Corresponding Authors**

*E-mail: jhzhao@semi.ac.cn; hzquantum@mail.tsinghua.edu.cn

**Notes**

The authors declare that they have no competing financial interests.

Data Availability: The data that support the transport plots within this paper and corresponding codes are available at https://doi.org/10.5281/zenodo.4292872

ACKNOWLEDGMENTS

D.P., H.D.S and S.Z. contributed equally to this work. This work was supported by the National Natural Science Foundation of China (Grant Nos. 61974138 and 11704364), Beijing Natural Science Foundation (Grant No. 1192017) and Tsinghua University Initiative Scientific Research Program. D.P. also acknowledges the support from Youth Innovation Promotion Association, Chinese Academy of Sciences (No. 2017156). H.D.S acknowledges China Postdoctoral Science Foundation (2020M670173 & 2020T130058).

**Figure captions**

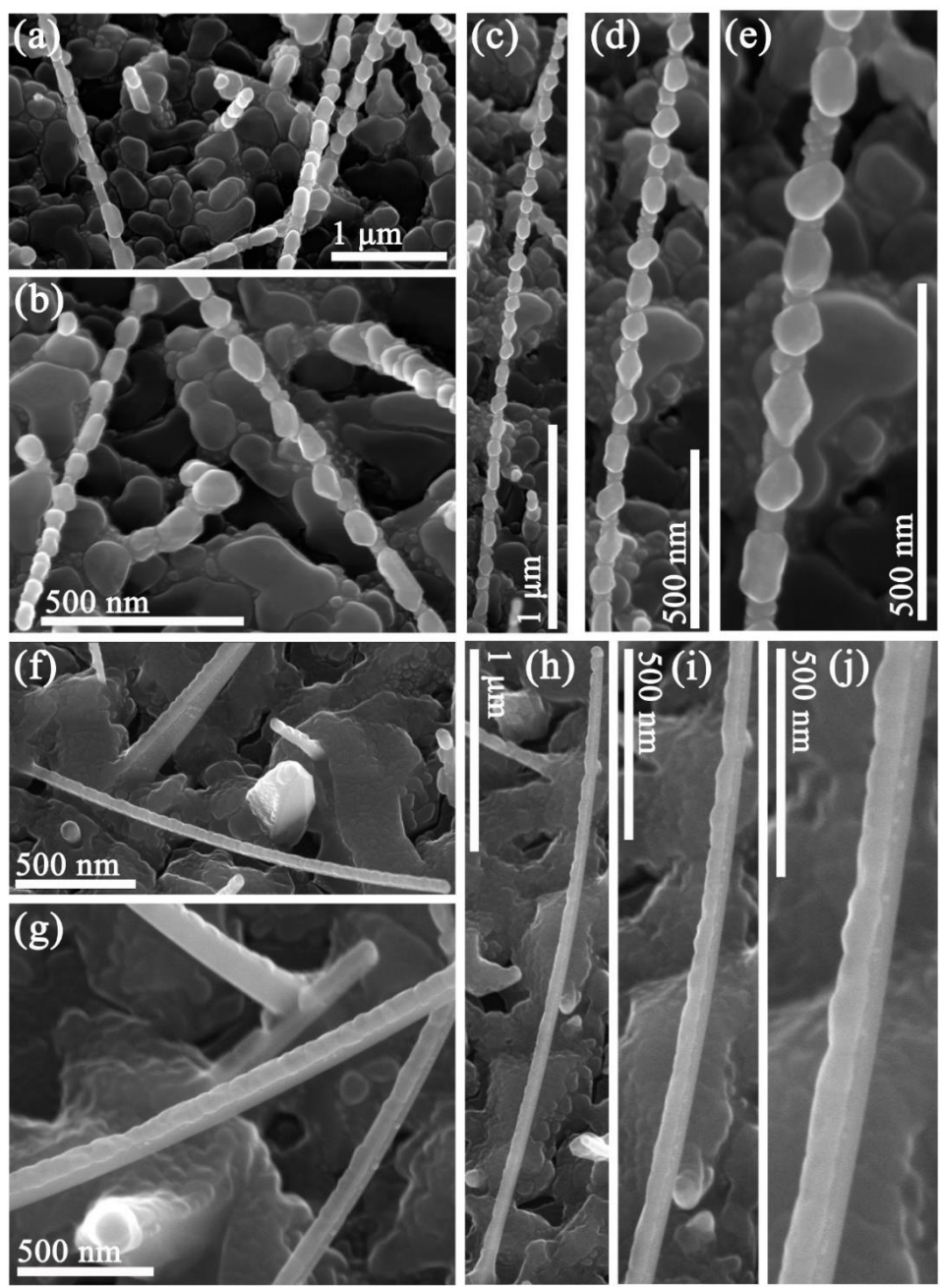

**Figure 1. Morphology control of the InAs-Al nanostructures.** a,b) SEM images of the InAs-Al recorded from different regions on the Si (111) substrates with the Al shell grown at ~ 1 °C. c-e) High magnification SEM images of a typical InAs-Al nanostructure (grown at ~1 °C). Al looks like 'pearls on a string'. f,g) SEM images of the InAs-Al nanowires recorded from different regions on the Si (111) substrates with the Al shell grown at ~ -10 °C. h-j) High magnification SEM images of a typical InAs-Al nanowire (grown at ~ -10 °C). Continuous half Al shells form on the InAs facets. All the SEM images were taken at a tilt angle of 25°.

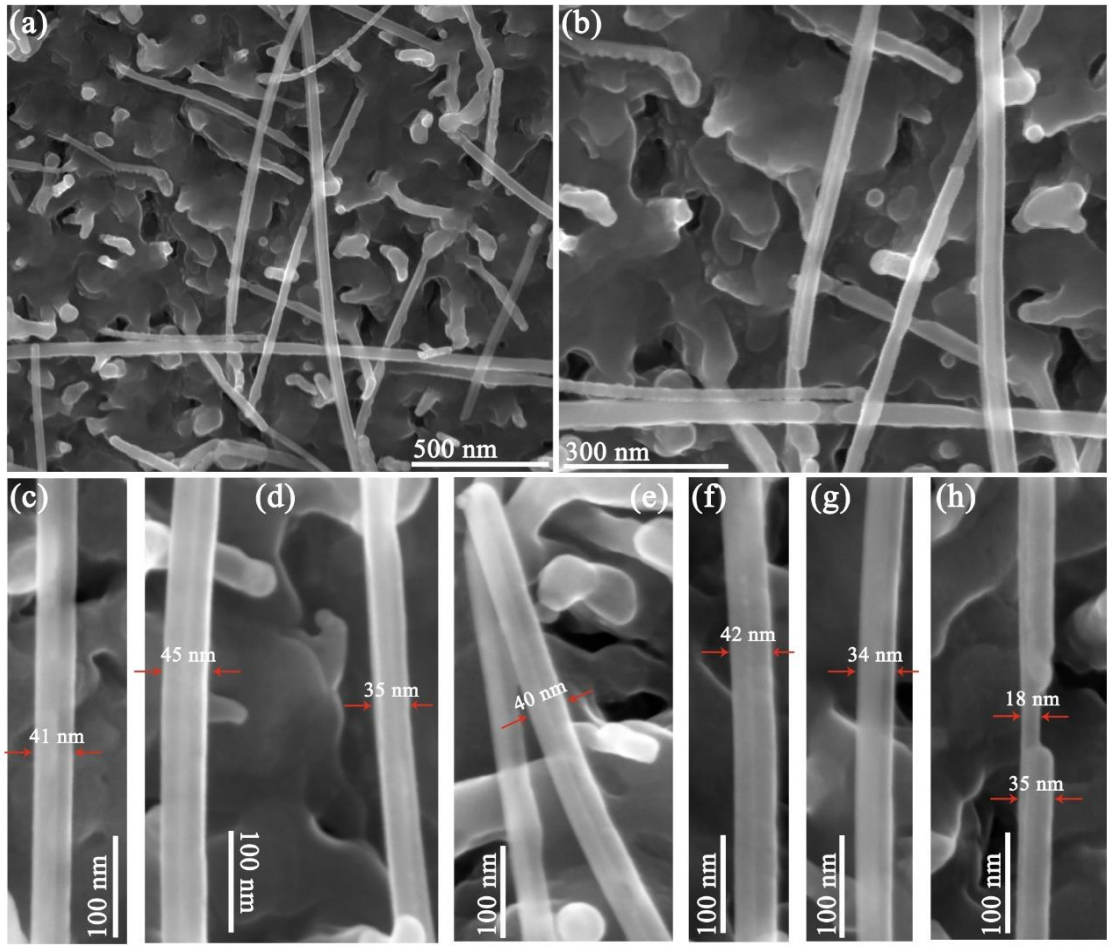

**Figure 2. Ultra-thin InAs-Al nanowires grown under the optimum conditions.** a) SEM image of the ultra-thin InAs-Al nanowires with the Al growth temperature and growth time at ~ -40 °C and 3 min, respectively. Continuous and smooth half Al shells form on the InAs facets. b) Enlarge view of (a). c-h) The close-up SEM images of the ultra-thin InAs-Al nanowires. All the SEM images taken at a tilt angle of 25°.

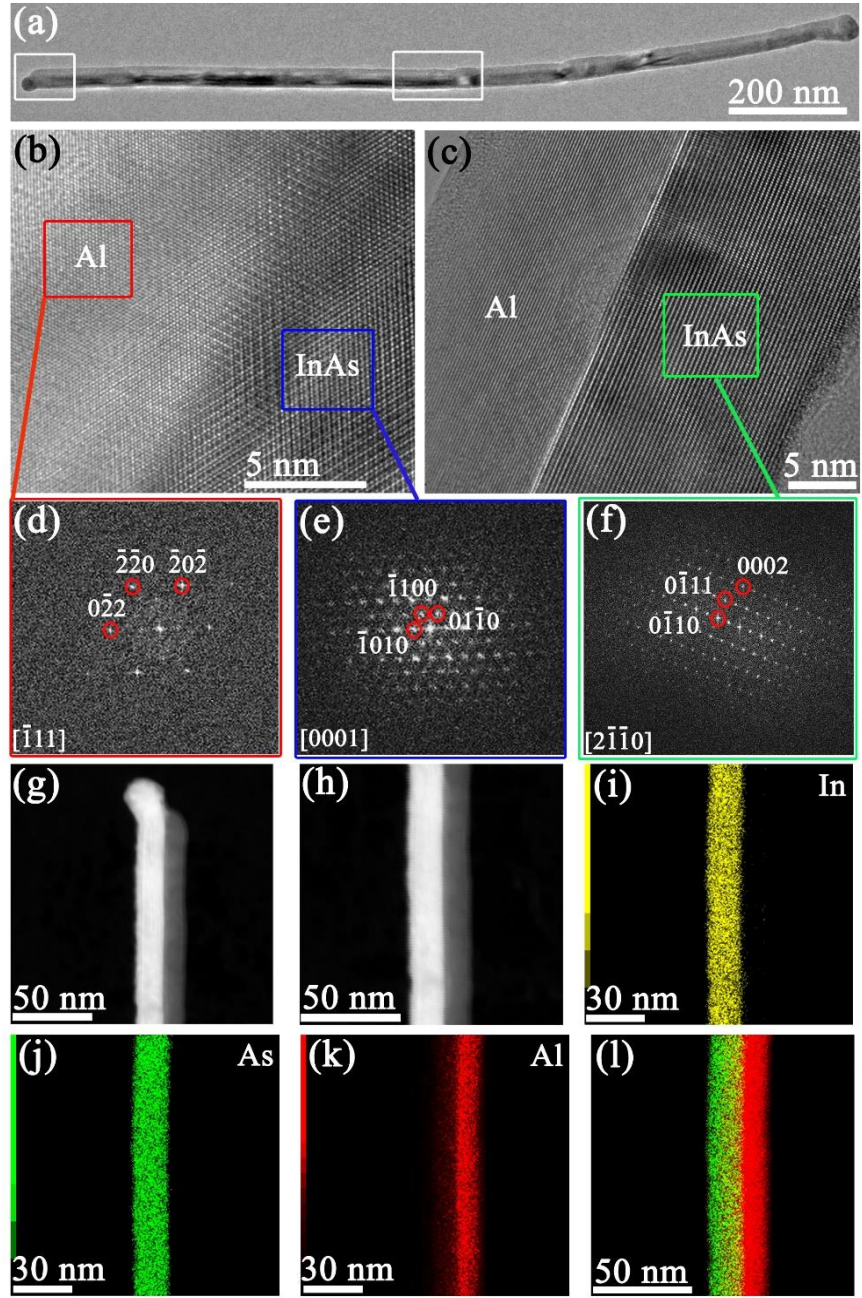

**Figure 3. Crystal structure and chemical composition of the ultra-thin InAs-Al nanowires**. a) Low-magnification TEM image of an InAs-Al nanowire. b,c) High-resolution TEM images of the InAs-Al nanowire viewed along the [0001] and [2-1-10] axes, respectively. d,e) The corresponding FFTs of the Al and InAs segments viewed along the [0001] axis. f) The corresponding FFT of the InAs segment viewed along the [2-1-10] axis. g,h) HAADF-STEM images taken from the top and middle sections of the InAs-Al nanowire, respectively. The white rectangles highlight the regions where HAADF-STEM images were recorded. i-k) False-color EDS elemental maps of In (yellow), As (blue), and Al (red) taken at the middle region of the InAs-Al nanowire, respectively. l) False-color overlay EDS elemental maps of In (yellow), As (blue), and Al (red).

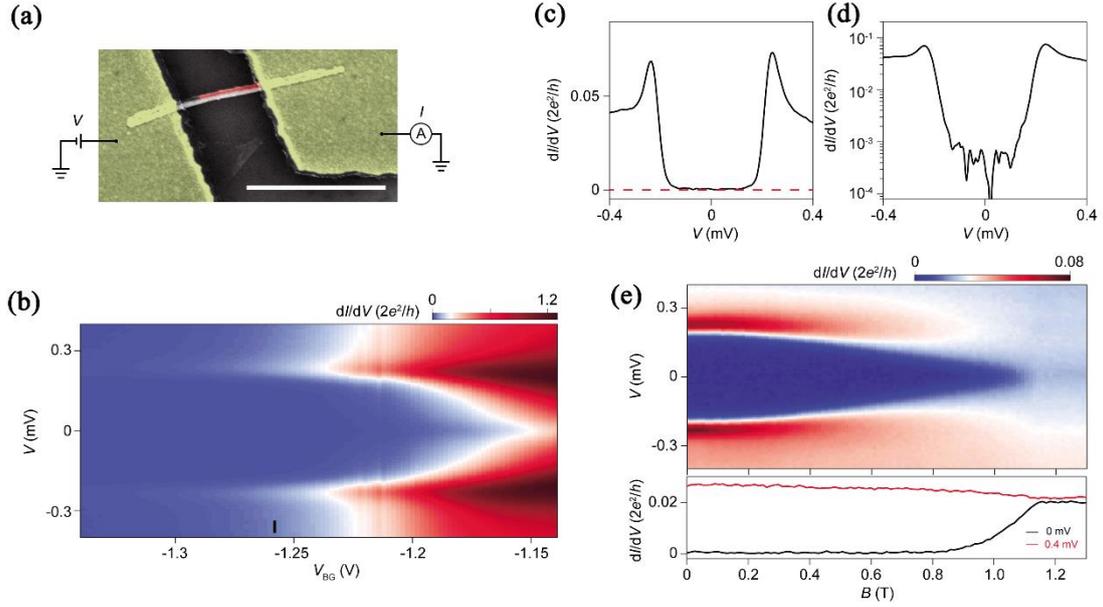

**Figure 4. Hard superconducting gap.** a) False-color SEM image of an N-NW-S device (Device A). Scale bar is 1 μm. Part of the Al film (red) on the ultra-thin InAs nanowire (grey) was selectively etched. The nanowire was then contacted by normal metal (yellow, 10 nm Ti and 70 nm Au). The substrate is p-doped Si, acting as a global back gate, covered by 285 nm thick $SiO_2$ (gate dielectric). Fridge base temperature is ~ 20 mK for all the measurement with various fridge filters. b) Differential conductance ($dI/dV$) of Device A as a function of bias voltage ($V$) and back gate voltage ($V_{BG}$) resolving the superconducting gap of $\Delta \sim 0.24$ meV. c) and d) Vertical line-cut (linear and log-scale) at $V_{BG}$ of -1.258 V (labeled by the black bar in panel (b) resolves a hard superconducting gap. e) $B$-dependence of the gap with $B$-direction aligned with the wire axis. Lower panel shows the horizontal line-cuts within the gap (0 bias, black curve) and outside the gap ($V = 0.4$ mV, red curve).

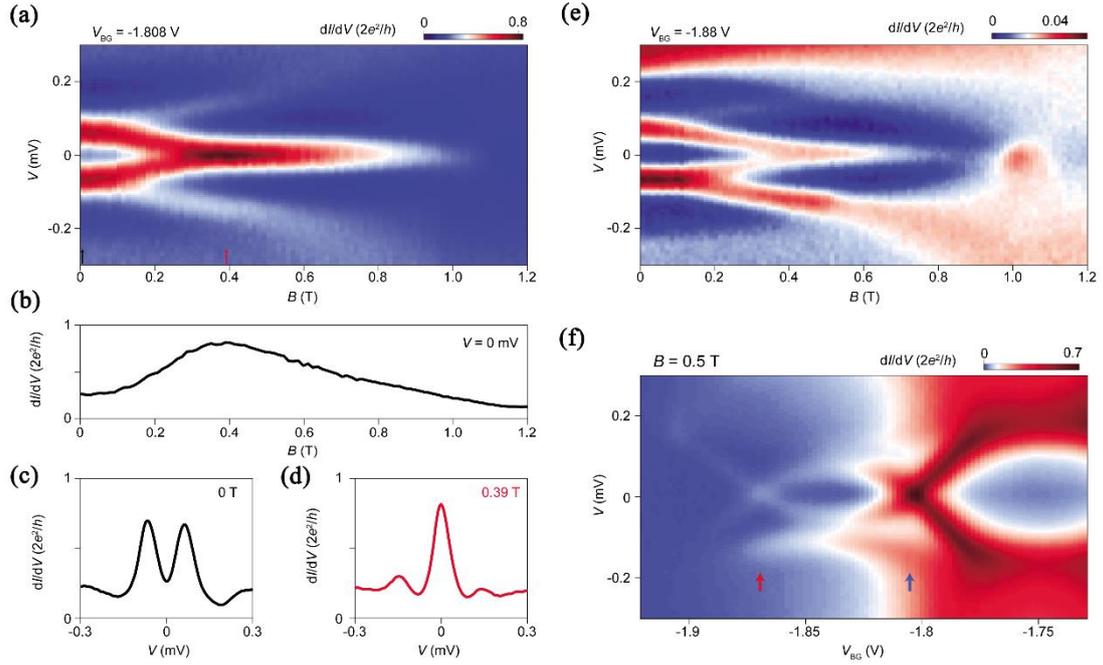

**Figure 5. Large zero bias peak.** a) d$I$/d$V$ of Device A shows a large zero bias peak (ZBP) at a $V_{BG}$ of -1.808 V. b) Zero-bias line-cut shows the peak height exceeding 80% of $2e^2/h$. c) and d) d$I$/d$V$ curve (vertical line-cut) at 0 T and 0.39 T, respectively. e) Another ZBP remains non-split against the *B*-sweep at a different $V_{BG}$ of -1.88 V. f) $V_{BG}$ sweep of the two ZBPs from (a) and (e) at $B$ = 0.5 T. The blue and red arrows roughly correspond to the gate voltages of (a) and (e), respectively, with a small gate sweep hysterisis.

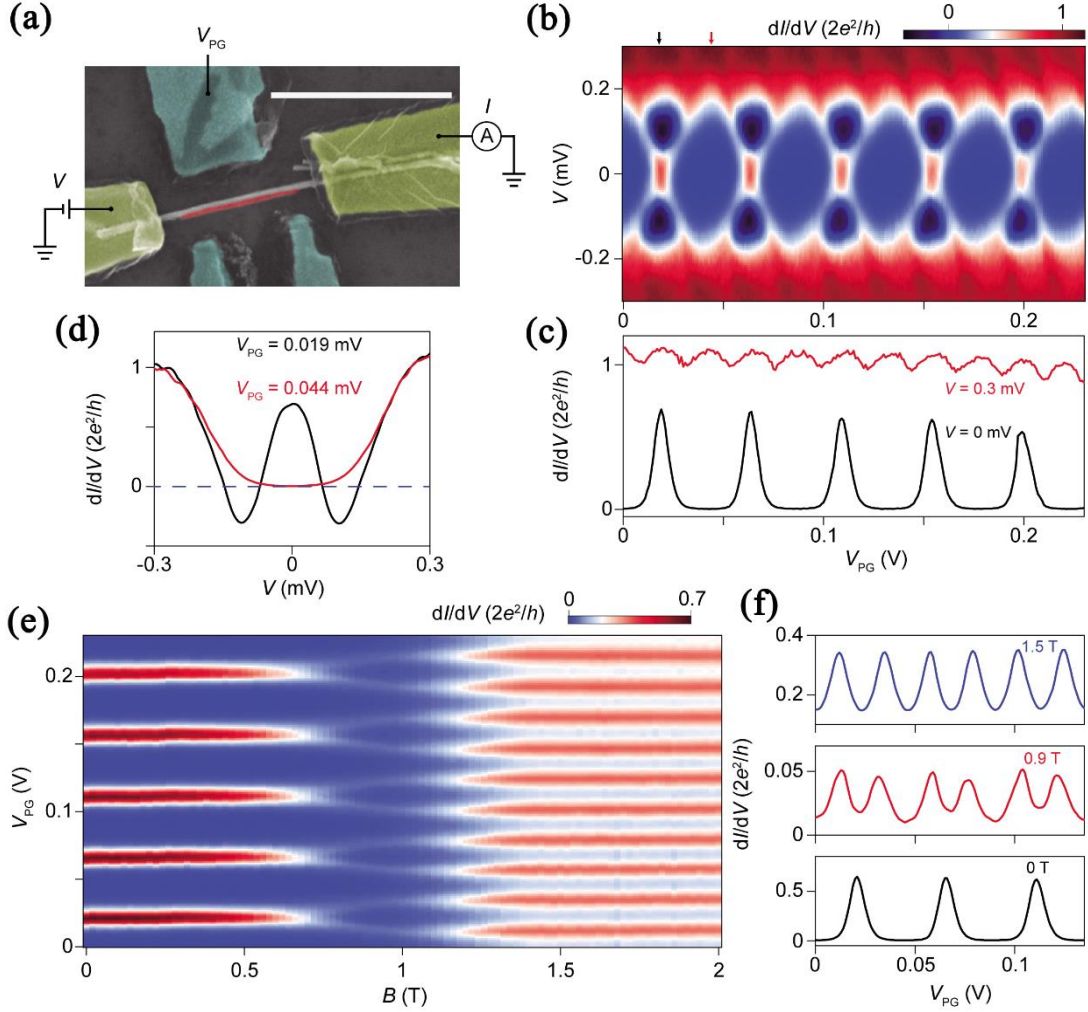

**Figure 6. 2*e*-periodic Coulomb blockade and 2*e*-1*e* transition of an island device.**
a) False-color SEM image of an InAs-Al island device (Device B). Scale bar is 1 μm. The Al island (red) was defined by wet chemical etching. The normal metal contact (yellow) and side gates (cyan) are Ti/Au (10 nm/70 nm). The plunger gate voltage ($V_{PG}$) tunes electron number on the island, while the lower two tunnel-gates tune the tunnel coupling between the island and the contacts. b) d$I$/d$V$ of Device B as a function of $V$ and $V_{PG}$ reveals regular Coulomb-blockade diamonds. The two tunnel gates and back gate were kept grounded for the measurement. c) Horizontal line-cuts at $V = 0$ mV (black curve) and $V = 0.3$ mV (red curve). d) Vertical line-cuts at the Coulomb valley (red curve) and Coulomb peak degeneracy point (black curve) indicated by the corresponding arrow in (b). e) d$I$/d$V$ (at $V = 0$) as a function of $V_{PG}$ and $B$ (direction along the nanowire) reveals the 2*e*-1*e* transition, with the vertical line-cuts at different $B$-values shown in (f).

# Supporting Information

# *In Situ* Epitaxy of Pure Phase Ultra-Thin InAs-Al Nanowires for Quantum Devices

Dong Pan,[†,§,⊥] Huading Song,[‡,§] Shan Zhang,[‡] Lei Liu,[†] Lianjun Wen,[†] Dunyuan Liao,[†] Ran Zhuo,[†] Zhichuan Wang,[△] Zitong Zhang,[‡] Shuai Yang,[‡,§] Jianghua Ying,[‡,§] Wentao Miao,[‡] Yongqing Li,[△] Runan Shang,[§] Hao Zhang,[*,‡,§,∥] and Jianhua Zhao[*,†,§,⊥]

[†]*State Key Laboratory of Superlattices and Microstructures, Institute of Semiconductors, Chinese Academy of Sciences, P.O. Box 912, Beijing 100083, China*
[‡]*State Key Laboratory of Low Dimensional Quantum Physics, Department of Physics, Tsinghua University, 100084 Beijing, China.*
[§]*Beijing Academy of Quantum Information Sciences, 100193 Beijing, China.*
[△]*Beijing National Laboratory for Condensed Matter Physics, Institute of Physics, Chinese Academy of Sciences, Beijing 100190, China*
[∥]*Frontier Science Center for Quantum Information, 100084 Beijing, China*
[⊥]*Center of Materials Science and Optoelectronics Engineering, University of Chinese Academy of Sciences, Beijing, 100190, China*

*To whom correspondence should be addressed. E-mails: jhzhao@semi.ac.cn (J.H.Z.); hzquantum@mail.tsinghua.edu.cn (H.Z.)

**Section 1 TEM and high-resolution TEM images of the InAs-Al nanowires with the Al shell grown at ~ -10 °C**

**Section 2 Parameters for continuous Al layer growth**

**Section 3 TEM and high-resolution TEM images of the ultra-thin InAs-Al nanowires grown under the optimum conditions**

**Section 4 ZBP phase diagram of Device A by fixing *B* and scanning gate**

**Section 5 ZBP phase diagram of Device A by fixing gate and scanning *B***

**Section 6 2*e*- Coulomb oscillations in a second island device**

**Section 1 TEM and high-resolution TEM images of the InAs-Al nanowires with the Al shell grown at ~ -10 °C**

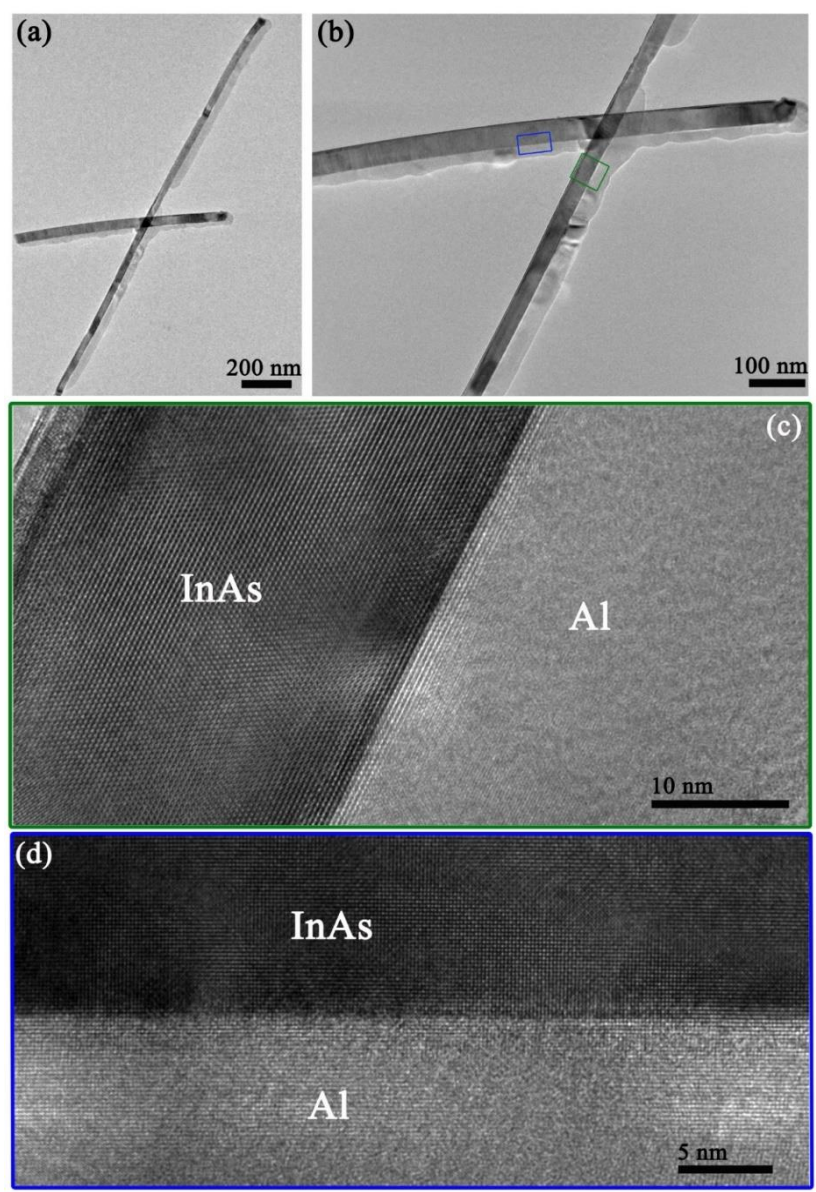

**Supplement Fig.S1** a,b) TEM images of InAs-Al nanowires with the Al shell grown at ~ -10 °C. c,d) High-resolution TEM images of the InAs-Al nanowires in (b) indicated by the corresponding color boxes (green and blue).

  **Figure S1** shows the TEM images of InAs-Al nanowires with the Al shell grown at ~ -10 °C. Continuous but rough and faceted outer surface of Al shells can be observed, consistent with the SEM results in **Figure 1**f-g in the main text. The interface between InAs nanowire and Al shell is sharp since the InAs nanowires are pure phase crystals, free of stacking faults and twin defects, which can provide smooth InAs side surfaces for Al growth (Figure S1c,d).

  As mentioned in the main text, one possible reason for the rough and faceted outer surface of Al shells could be that the -10 °C substrate temperature is still not low enough and the Al growth follows the high temperature dynamics. That is, well-separated large Al grains were initially nucleated during the growth with different preferred crystal orientations appeared at a later stage, forming the grain boundaries and strains. Thus, as shown in **Figure S2**, polycrystalline Al shells which consist of type-α and -β grains can be observed by TEM. This phenomenon has also been observed in previous work[1,2].

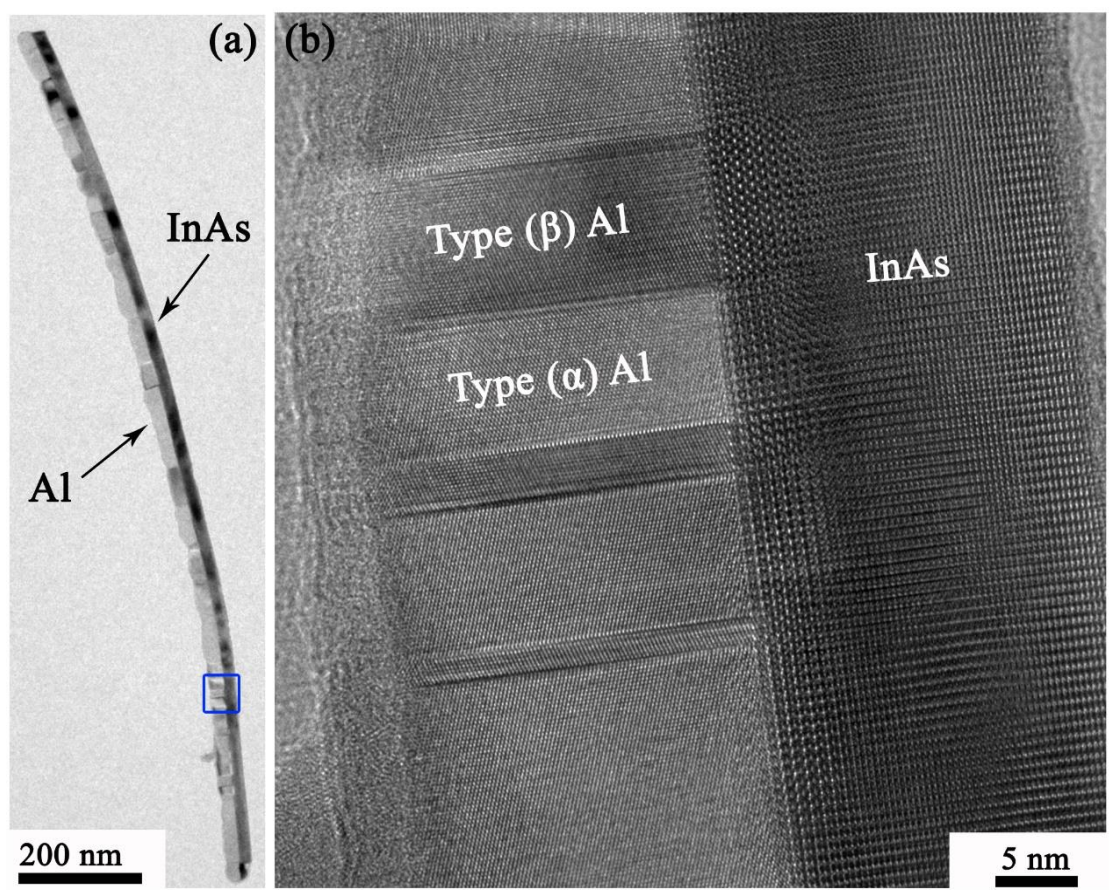

**Supplement Fig.S2** a) TEM image of an InAs-Al nanowire with the Al shell grown at ~ -10 °C. b) High-resolution TEM image of the nanowire in (a) (blue box). Polycrystalline Al shell which consists of type-α and -β grains can be observed.

**Section 2 Parameters for continuous Al layer growth**

In our work, all the InAs-Al nanowires were grown on p-type Si (111) substrates by MBE using Ag as the catalysts. From the SEM images in the main text (Figure 1 and 2), we can see that most InAs nanowires have non-<111> growth directions. The possible reasons for this phenomenon are as follows. It is known that [111] (or [0001]) is the polar direction of III-V semiconductor nanowires with zinc-blende (or wurtzite) structure. Group-III and group-V terminated surfaces are defined as (111)A and (111)B polar surfaces, respectively. However, the non-polar nature of Si substrates indicates that (111)A and (111)B polar surfaces can coexist on the Si substrates surface, which normally leads to random growth directions of III-V nanowires. Meanwhile, we know that the Si substrates are very easy to oxidize in air. As mentioned in the experimental section, the Si substrates used in this work were immersed in a diluted HF (2%) solution for 1 min to remove the surface contamination and native oxide, but a thin oxide layer can form on Si substrates surface within a very short time before loading the Si substrates into the MBE chamber. According to literatures[3,4], Si substrates are usually annealed to about 900 °C under UHV conditions before MBE growth to completely remove the thin oxide layer before the growth process. However, this Si surface treatment temperature (~ 900 °C) exceeds the upper limit of the temperature (~ 800 °C) of our substrate heater. That means that the thin natural oxide layer of Si substrates cannot be removed completely before the InAs nanowire growth, which can result in random growth directions of our nanowires. Based on the literatures, the Al shell can be epitaxially grown either on the facets of <111> oriented InAs nanowires[1,5] or <001> oriented InAs nanowires[2]. Our work shows that, besides <111> and <001> orientations, the Al shells can also be epitaxially grown on the facets of InAs nanowires with various non-<111> growth directions due to the random growth directions of our nanowires.

We find that the substrate temperature plays a key role in the evolution of the Al shell morphology: a low substrate temperature is a prerequisite in enabling the deposition of continuous Al layers on the nanowire facets. **Figure S3** shows SEM images of the InAs-Al nanowires with the Al shell grown at ~ -40 °C. From the high magnification SEM image of the nanowires (Figure S3a,b), we can see that continuous, uniform and thin half Al shells can be successfully grown for almost all growth directions. For some rare cases, discontinuous Al shells have been found in the samples grown at different temperatures (ranging from ~ 1 °C to ~ -40 °C). As shown in **Figure S4**, this discontinuous and rough Al surface is usually accompanied by an obvious enlarged catalyst particle on the top of the InAs-Al nanowire, which can be easily identified by SEM. **Figure S5** shows the TEM analysis of such a case. HAADF-STEM image taken from the top of the wire indicates that the catalyst particle has two parts (Figure S5e). We can see from the false-color EDS elemental maps that the chemical compositions of the two parts are pure Al and Ag-In alloy, respectively. The growth mechanism of this kind of InAs-Al nanowires is unclear and it can be studied for future research. Besides growth temperature, the Al flux is also an important factor for the continuous Al layer growth. We find that for Al grown at ~1°C with a very small Al flux, nano-size Al droplets can form on the InAs nanowire side walls. A higher beam flux and a lower temperature is beneficial for the continuous Al layer growth. In the main text, all the InAs-Al nanowires are grown with a high Al flux (~ $1.1 \times 10^7$ mbar).

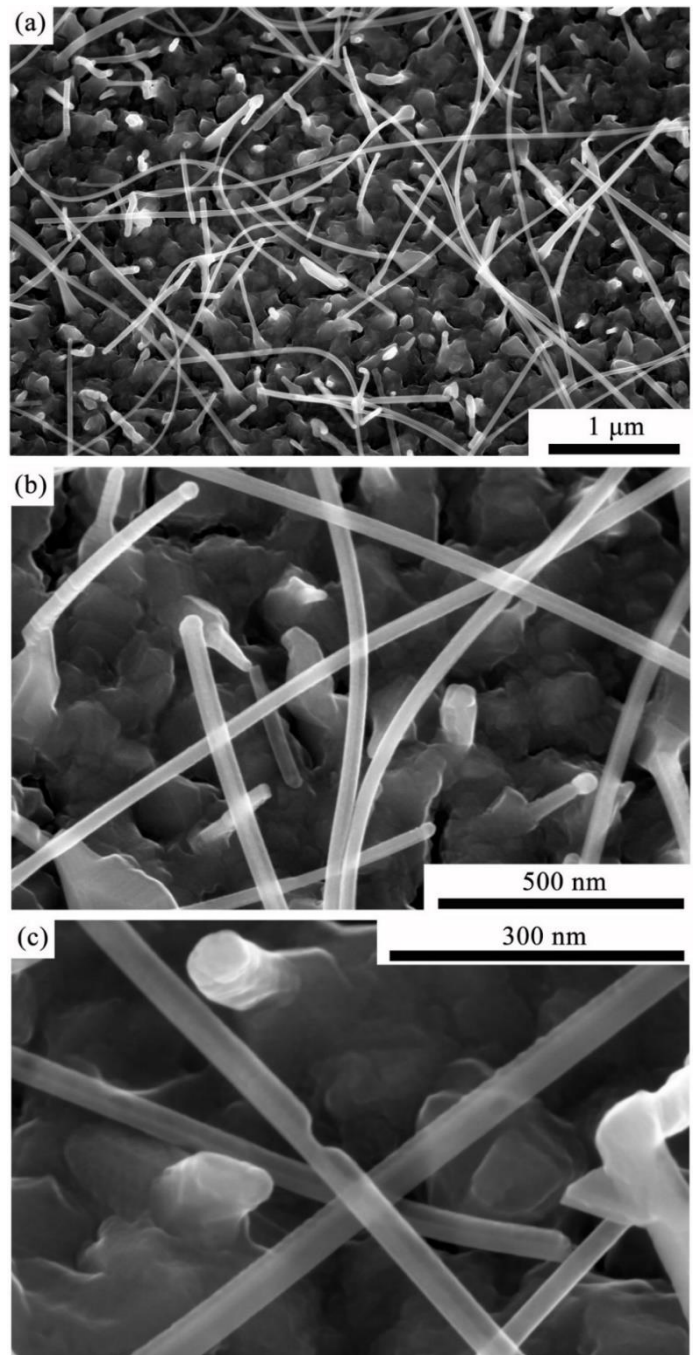

**Supplement Fig.S3** a) SEM image of the ultra-thin InAs-Al nanowires grown on Si (111) substrates with the Al shell grown at ~ -40 °C. b,c) High magnification SEM images of the ultra-thin InAs-Al nanowires taken from the same sample with (a). All the SEM images were taken at a tilt angle of 25°.

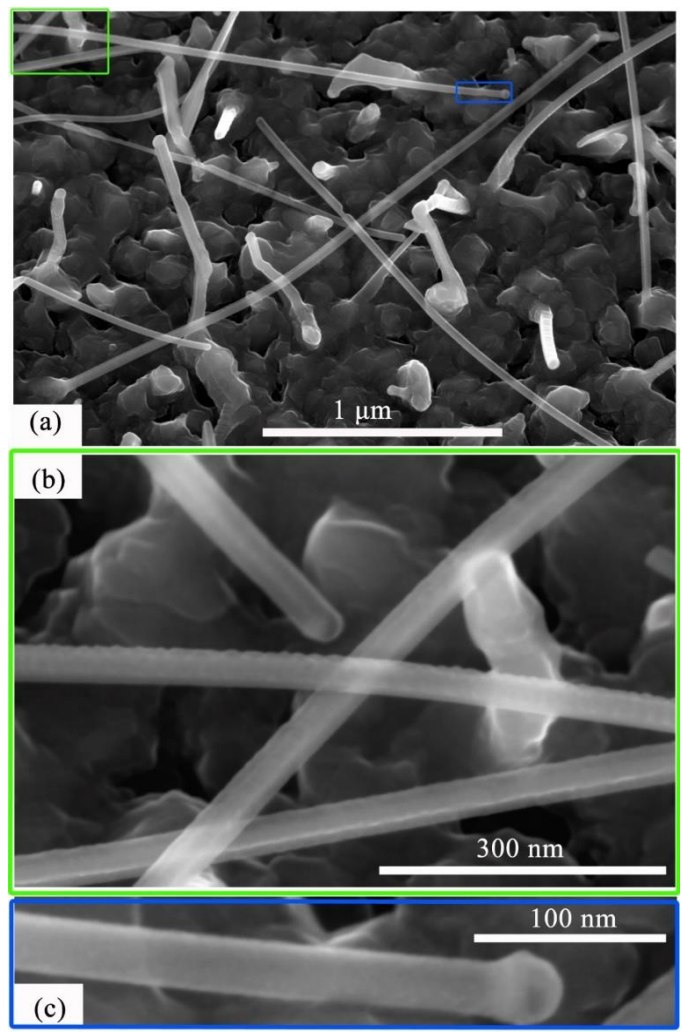

**Supplement Fig.S4** a) SEM image of ultra-thin InAs-Al nanowires with the Al shell grown at ~ -40 °C. An InAs-Al nanowire with an obvious enlarged catalyst particle can be observed (blue box). b,c) High magnification SEM images of this nanowire indicated by the corresponding color boxes in a). All the SEM images taken at a tilt angle of 25°.

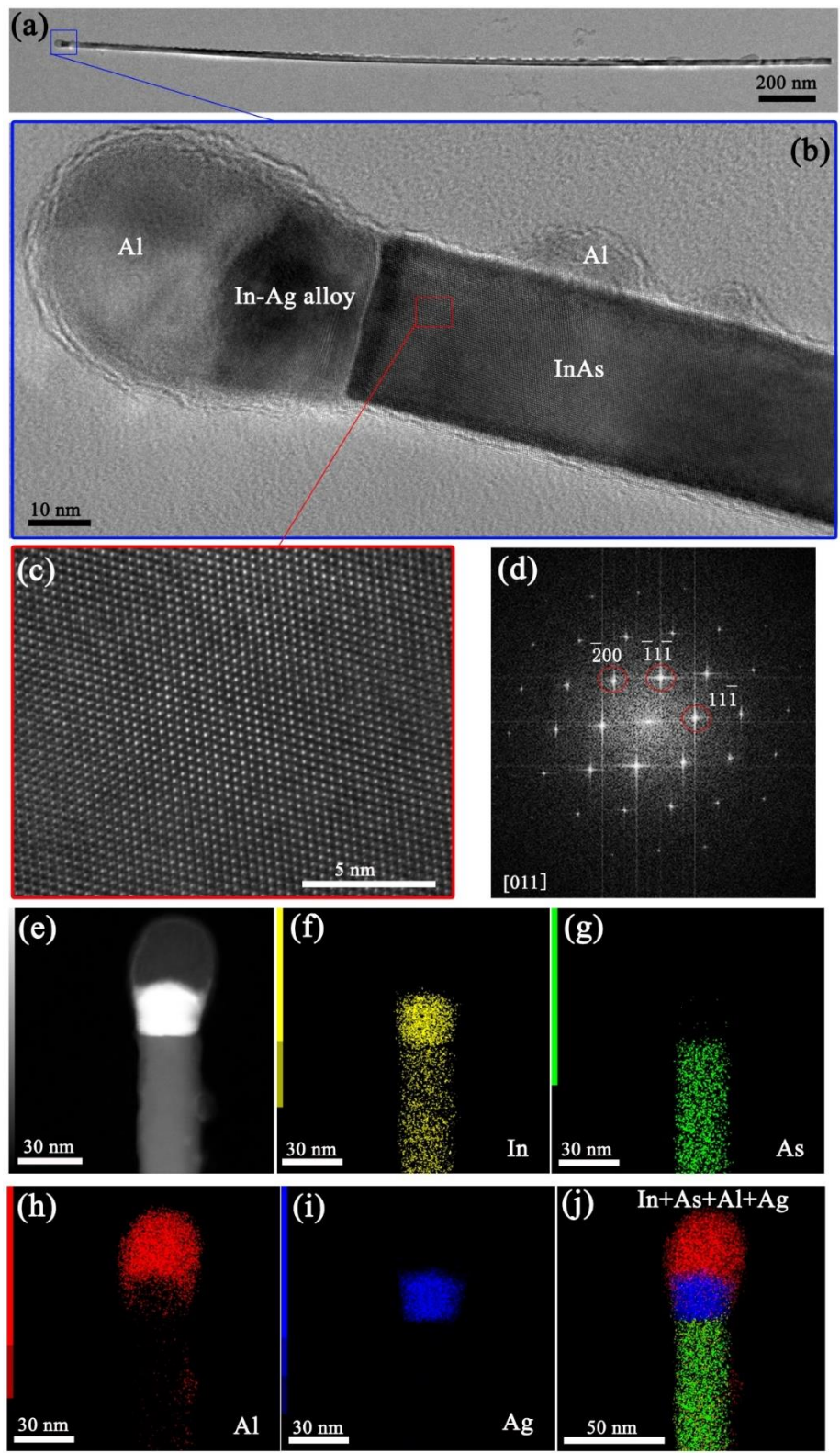

**Supplement Fig.S5** a-c) TEM and high-resolution TEM images of an ultra-thin InAs-Al nanowire with an obvious enlarged catalyst particle. d) FFT of (c) shows that the InAs nanowire is a pure ZB crystal. e) HAADF-STEM image taken from the top region of the InAs-Al nanowire. f-j) False-color EDS elemental maps of (e).

**Section 3 TEM and high-resolution TEM images of the ultra-thin InAs-Al nanowires grown under the optimum conditions**

Our work shows that high quality Al shell can be epitaxially grown on the facets of InAs nanowires with all the growth directions. **Figure S6-S11** show continuous and smooth Al half-shells formed on the facets of ultra-thin InAs nanowires grown along different directions: <21-1> (Figure S6), <-100> (Figure S7-S9), <13-1> (Figure S10), <0001> (Figure S11a) and <01-1> (Figure S11b). All the wires revealed an atomic scale uniformity of the InAs-Al interface and a perfect crystal structure, free of stacking faults. It is worth noting that high quality half Al shells can be epitaxially grown on both the pure WZ (Figure S6 and S11a) and pure ZB (Figure S7, S8, S9, S10 and S11b) structured InAs nanowires.

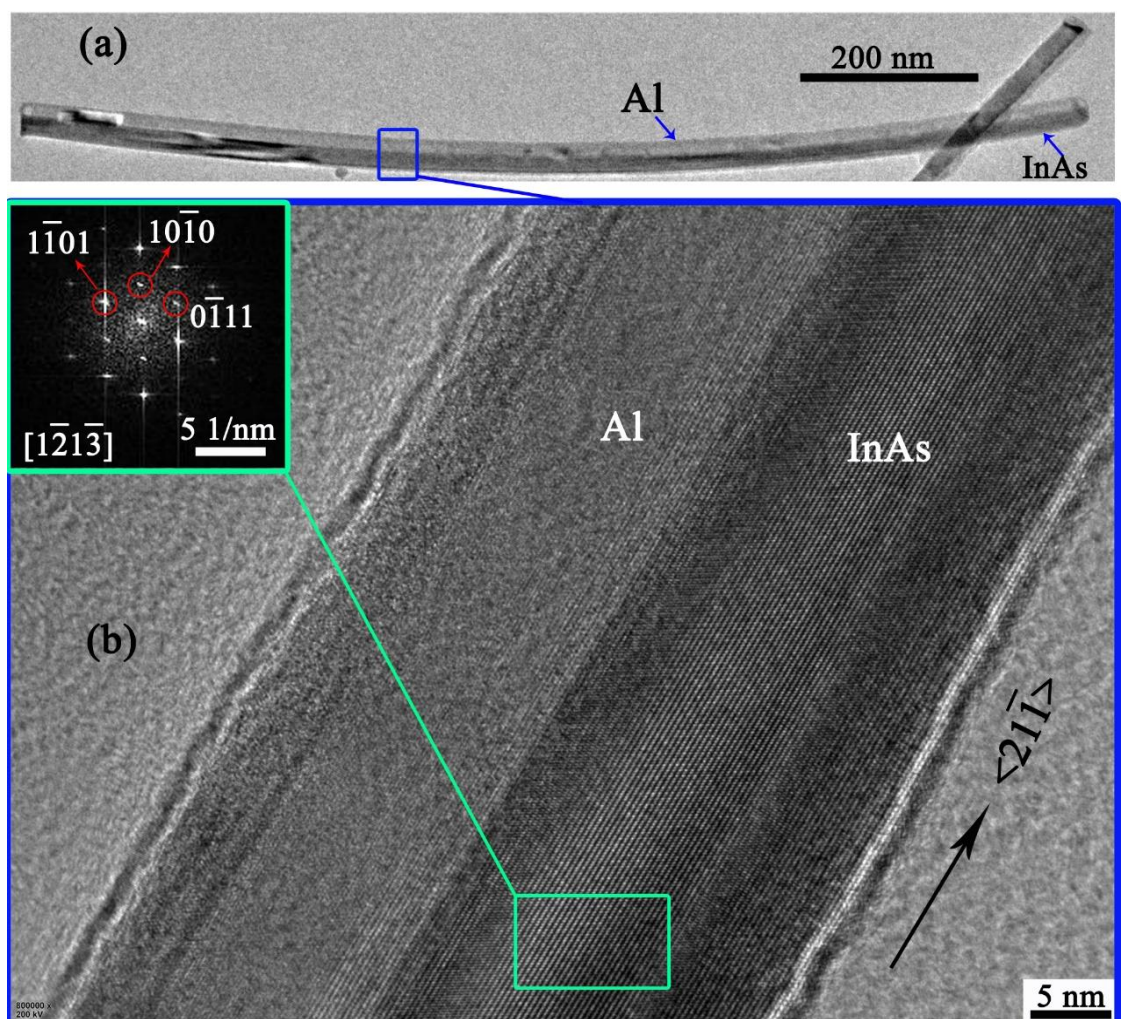

**Supplement Fig.S6** a) TEM image of an ultra-thin InAs-Al nanowire with the Al shell grown at ~ -40 °C. b) High-resolution TEM image of the blue box in (a). The inset is the corresponding FFT of the InAs in (b) (green box). Continuous and uniform half Al shell is epitaxially grown on the pure WZ structured ultra-thin InAs nanowire with growth along <21-1> direction.

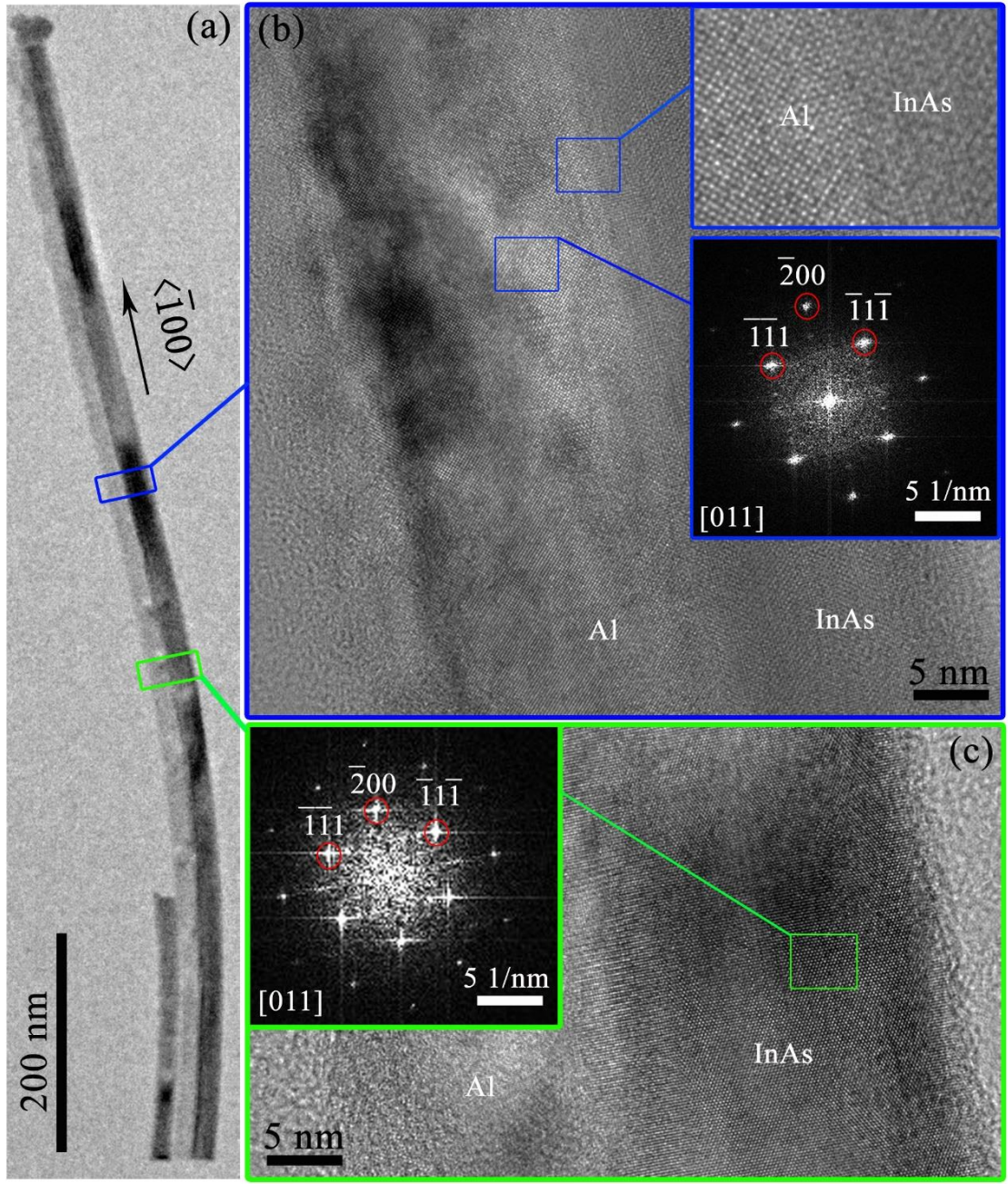

**Supplement Fig.S7** a) TEM image of an ultra-thin InAs-Al nanowire with the Al shell grown at ~ -40 °C. b,c) High-resolution TEM images recorded from the blue and green boxes in (a), respectively. The insets in (b) show the InAs-Al interface and the FFT of Al film. The inset in (c) is the FFT of the InAs part. Continuous and uniform half Al shell is epitaxially grown on the pure ZB structured ultra-thin InAs nanowire with growth along <-100> direction.

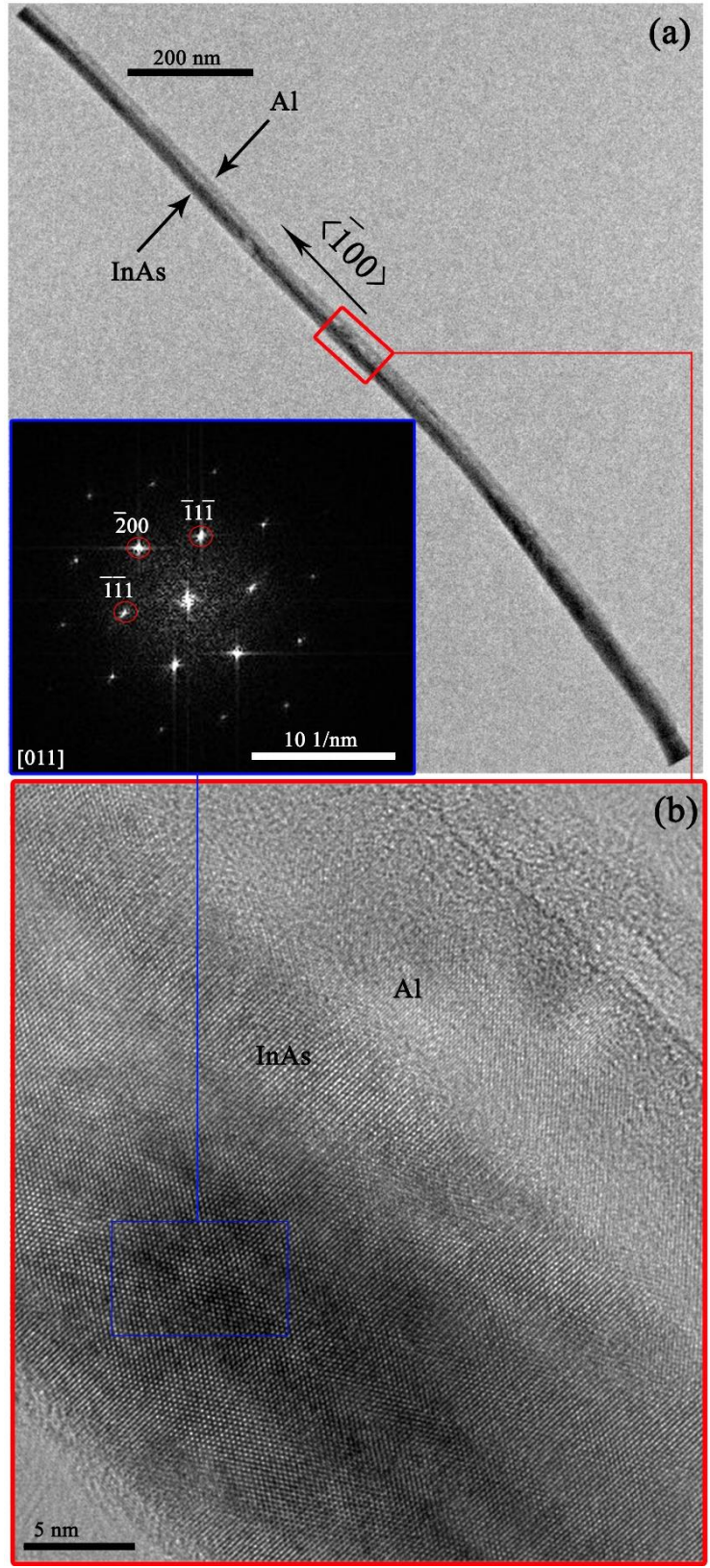

**Supplement Fig.S8** a) TEM image of an ultra-thin InAs-Al nanowire with the Al shell grown at ~ -40 °C and the nanowire grown along <-100> direction. b) High-resolution TEM image of the red box in (a). The inset is the corresponding FFT pattern of the InAs in (b). Pure ZB structured ultra-thin InAs-Al nanowire is obtained.

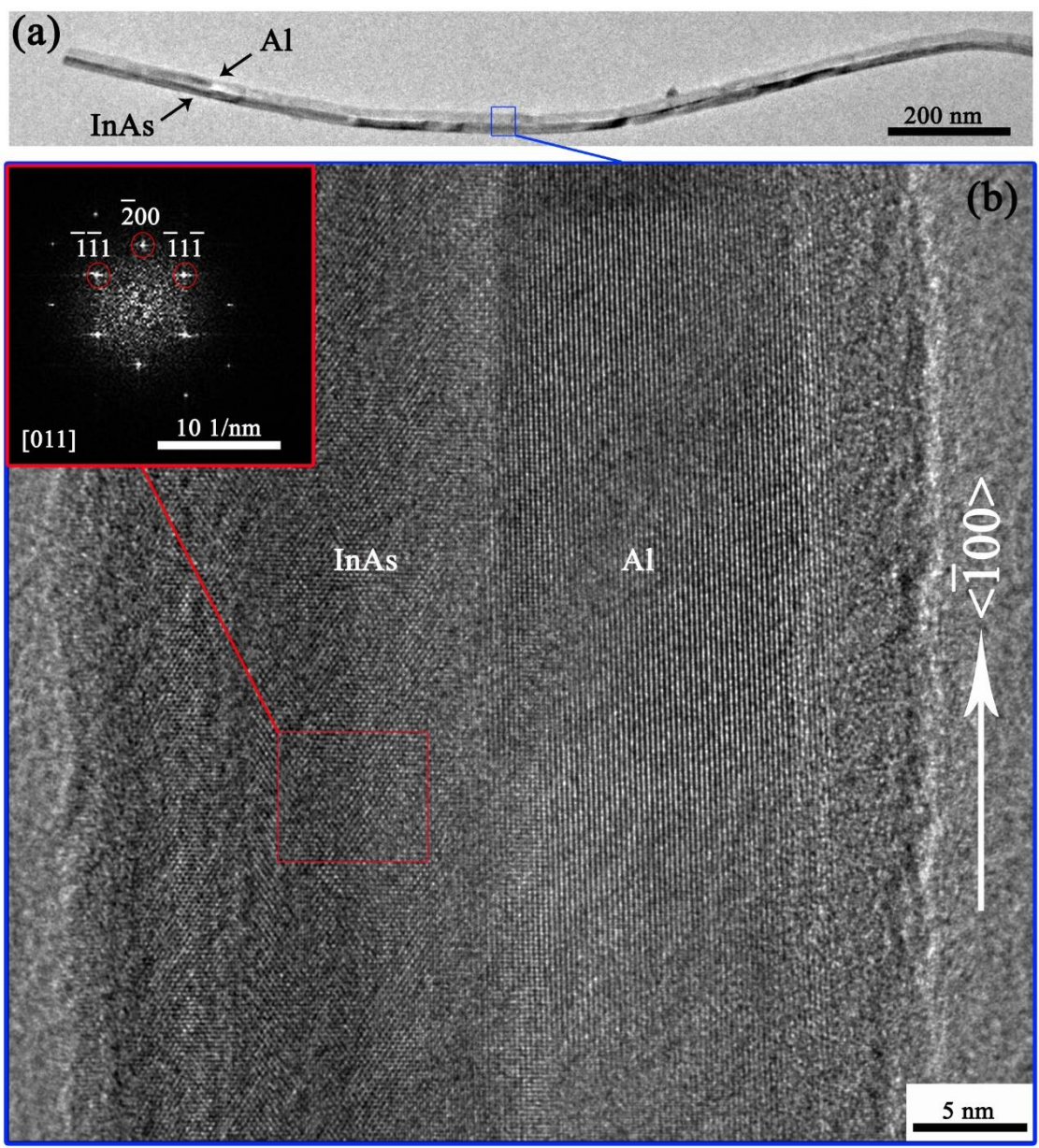

**Supplement Fig.S9** a) TEM image of an ultra-thin InAs-Al nanowire with the Al shell grown at ~ -40 °C and the nanowire grown along <-100> direction. b) High-resolution TEM image of the InAs-Al nanowire in (a). The inset is the corresponding FFT pattern of the InAs in (b). Pure ZB structured ultra-thin InAs-Al nanowire is obtained.

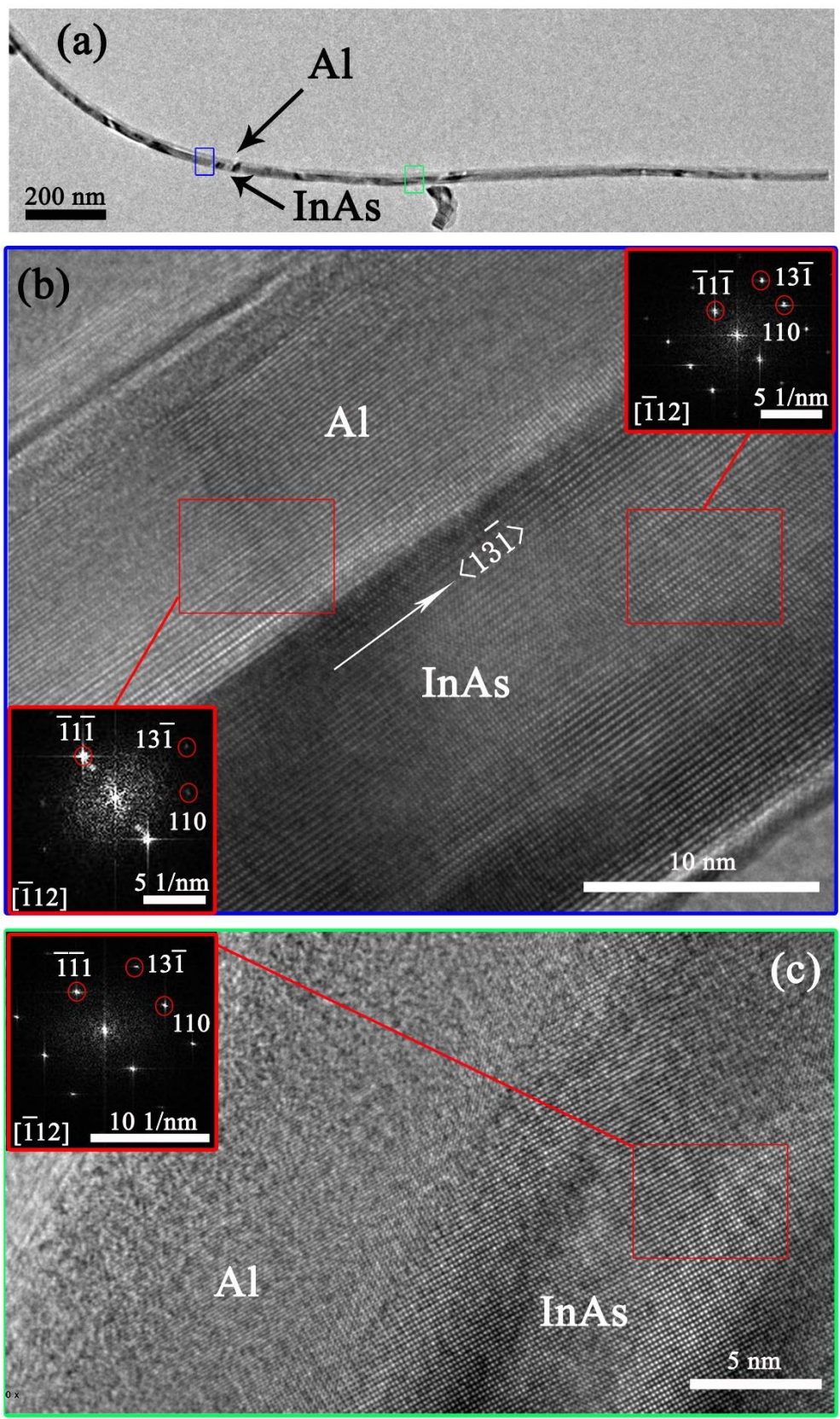

**Supplement Fig.S10** a) TEM image of an ultra-thin InAs-Al nanowire with the Al shell grown at ~ -40 °C and the nanowire grown along <13-1> direction. b,c) High-resolution TEM images of the InAs-Al nanowire recorded from the blue and green rectangles in (a), respectively. The insets in (b) and (c) are the FFT patterns of the corresponding high-resolution TEM images.

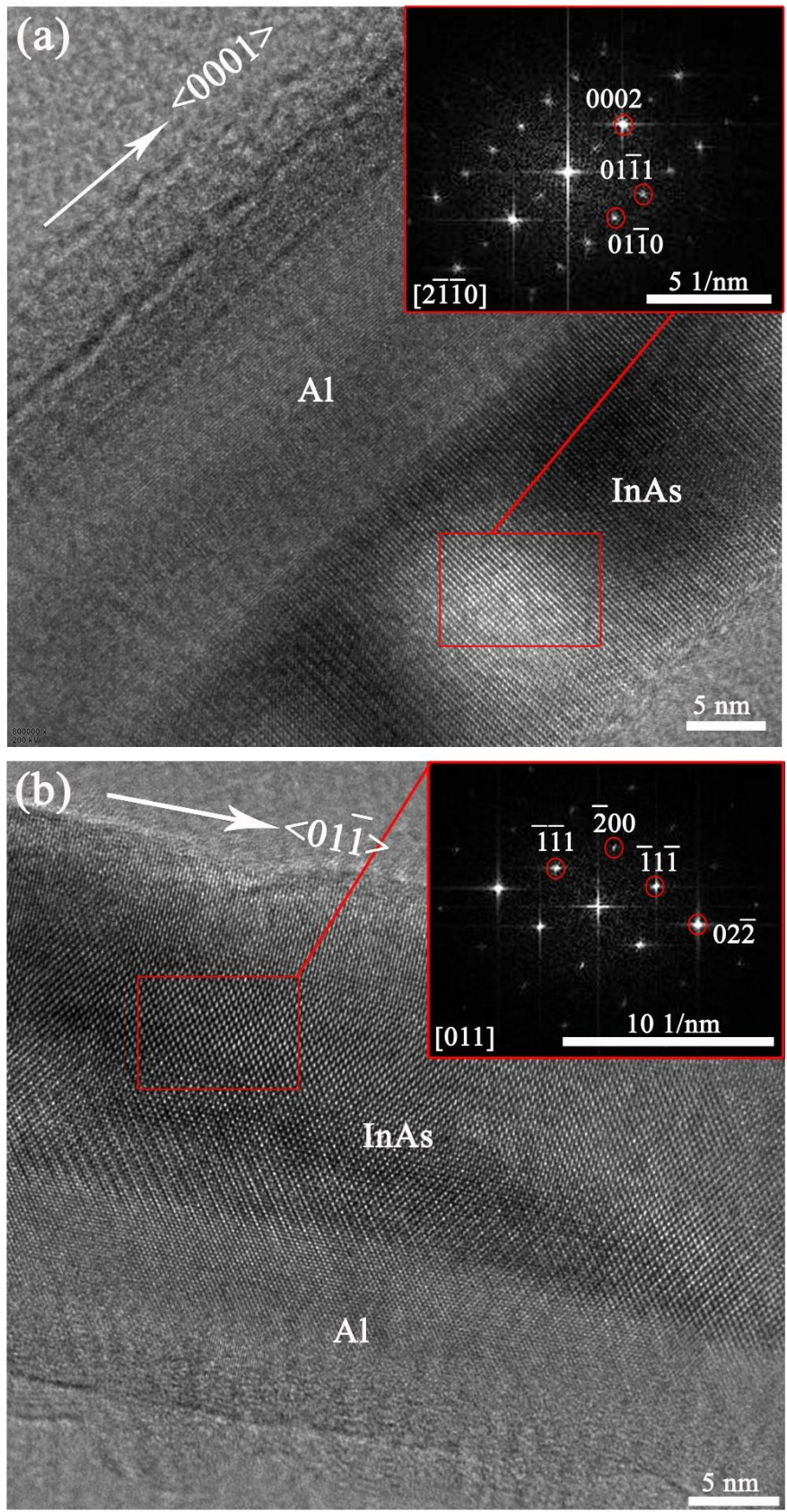

**Supplement Fig.S11** High-resolution TEM images of the ultra-thin InAs-Al nanowires with the Al shell grown at ~ -40 °C and along <0001> (a) and <01-1> (b) directions, respectively. The insets in (a) and (b) are the FFT patterns of the corresponding high-resolution TEM images. InAs nanowire in (a) has a pure WZ crystal structure and the nanowire in (b) has a pure ZB crystal structure.

## Section 4 ZBP phase diagram of Device A by fixing *B* and scanning gate

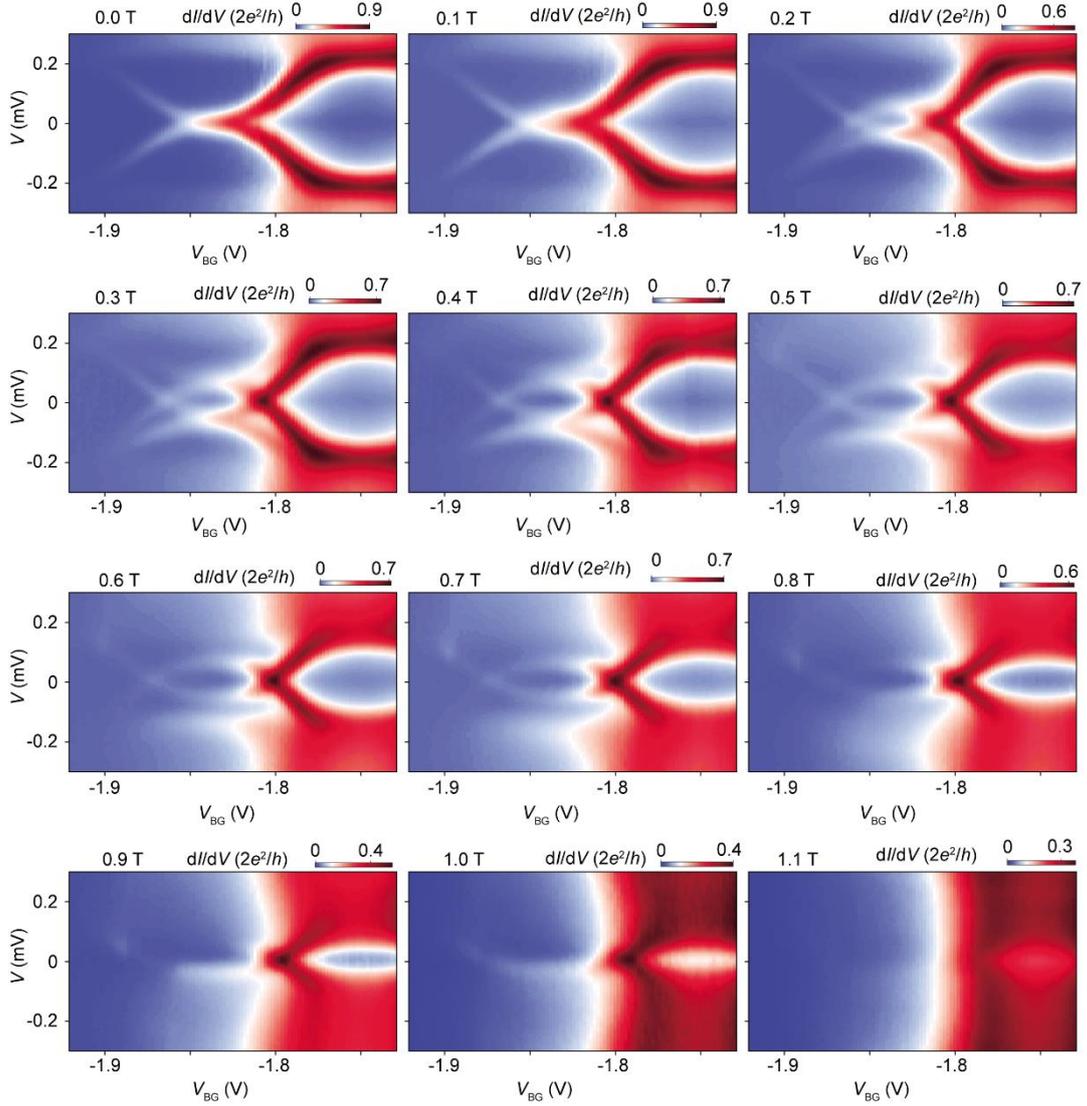

**Supplement Fig.S12 ZBP phase diagram of Device A by fixing *B* and scanning gate.** Each panel represents the $V_{BG}$ dependence of the Andreev bound state at a fixed *B*-field (labeled on the top left, from 0 T to 1.1 T with 0.1 T step). These panels show a continuous evolution of the Andreev bound state. The zero bias peak at finite *B* is due to the Zeeman splitting of the Andreev peak and the peak crossing. The $V_{BG}$ values corresponding to the two crossing points barely change at different *B*-fields. Therefore, by fine-tuning $V_{BG}$ to these two values, we can observe the 'robust' non-split ZBPs in Fig. 5a and Fig. 5e.

**Section 5 ZBP phase diagram of Device A by fixing gate and scanning *B***

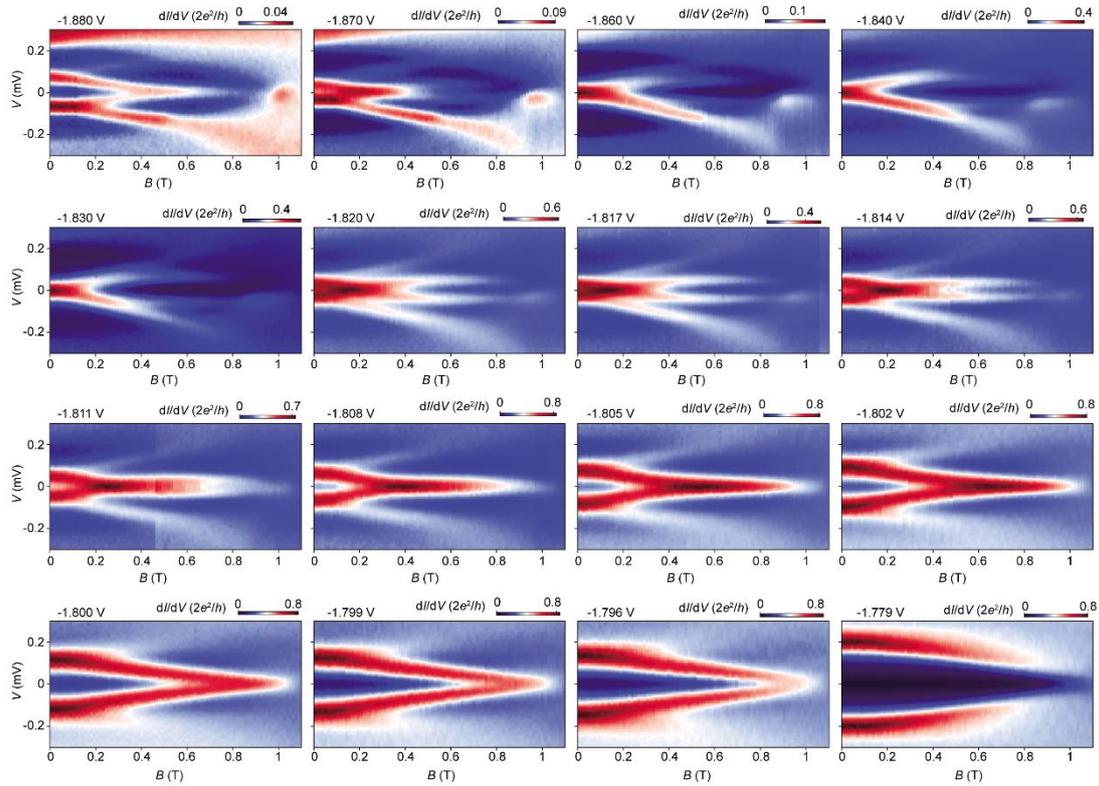

**Supplement Fig. S13 ZBP phase diagram of Device A by fixing gate and scanning *B*.** Each panel represents the *B*-dependence of the Andreev bound state at a fixed $V_{BG}$ (labeled on the top left, from -1.88 V to -1.779 V). These panels show a continuous evolution from one fine-tuned ZBP (-1.88 V) to the other one (-1.808 V), and finally no ZBP observed (-1.779 V) during *B*-sweep.

# Section 6 2*e*- Coulomb oscillations in a second island device

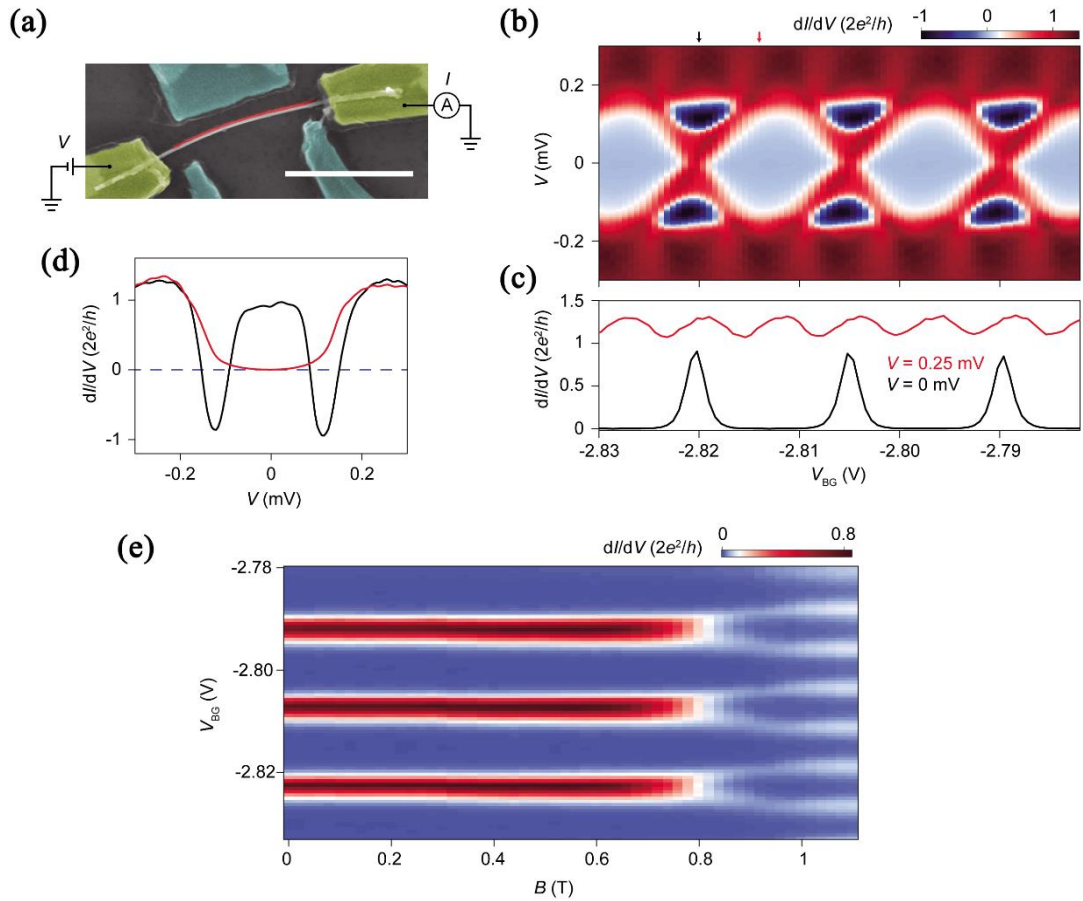

**Supplement Fig. S14 A second island device (Device C) showing 2*e*- Coulomb oscillations.** a) Device SEM (false-color) and measurement set up. Scale bar is 1 μm. b) d*I*/d*V* as a function of bias (*V*) and back gate voltage ($V_{BG}$) resolves the 2*e*- Coulomb diamond. The side gates were kept grounded during the measurement. The diamond size (*V* ~ 0.25 mV) corresponds to a charging energy of ~ 31 μeV. c) Horizontal line-cuts at *V* = 0 mV (black) and *V* = 0.25 mV (red) resolves the 2*e*- (Cooper pair) and 1*e*- (quasi-particle) oscillations. d) Vertical line-cuts at Coulomb valley (red) and Coulomb peak degeneracy point (black) with $V_{BG}$ labeled by the arrows in panel (b) with corresponding colors. The onset bias (~ 88 μeV) of the negative differential conductance gives an estimation of $E_O$ ~ 75 μeV. e) d*I*/d*V* at zero bias as a function of $V_{BG}$ and *B* (direction along the nanowire) shows a 2*e*-periodic to 1*e*-periodic transition.